\documentclass{article}
\usepackage{amsmath}
\usepackage{amssymb}
\usepackage{amsthm}

\def\eps{\varepsilon}
\def\reals{\mathbb{R}}

\def\uball{\mathbb{B}}

\def\comp{\raise 1pt \hbox{$\scriptstyle\circ$}}

\def\minimize{\mathop{\rm minimize}\limits}
\def\maximize{\mathop{\rm maximize}\limits}

\def\essinf{\mathop{\rm ess\ inf}\nolimits}
\def\st{\mathop{\rm subject\ to}}

\def\upto{{\raise 1pt \hbox{$\scriptstyle \,\nearrow\,$}}}
\def\downto{{\raise 1pt \hbox{$\scriptstyle \,\searrow\,$}}}
\def\inte{\mathop{\rm int}}
\def\cl{\mathop{\rm cl}}
\def\co{\mathop{\rm co}}
\def\lsc{\mathop{\rm lsc}}
\def\epi{\mathop{\rm epi}}

\def\tos{\rightrightarrows}
\def\FF{(\F_t)_{t=0}^T}

\def\B{{\cal B}}
\def\C{{\cal C}}
\def\D{{\cal D}}

\def\F{{\cal F}}

\def\M{{\cal M}}
\def\N{{\cal N}}

\newtheorem{theorem}{Theorem}
\newtheorem{lemma}[theorem]{Lemma}
\newtheorem{corollary}[theorem]{Corollary}

\newtheorem{example}{Example}
\newtheorem{remark}{Remark}
\theoremstyle{definition}

\title{Convex duality in stochastic programming and mathematical finance}
\author{Teemu Pennanen\footnote{Institute of Mathematics, Aalto University, P.O.\ Box 11100, FI-00076 Aalto, Finland, teemu.pennanen@tkk.fi}}

\begin{document}
\maketitle

\begin{abstract}

This paper proposes a general duality framework for the problem of minimizing a convex integral functional over a space of stochastic processes adapted to a given filtration. The framework unifies many well-known duality frameworks from operations research and mathematical finance. The unification allows the extension of some useful techniques from these two fields to a much wider class of problems. In particular, combining certain finite-dimensional techniques from convex analysis with measure theoretic techniques from mathematical finance, we are able to close the duality gap in some situations where traditional topological arguments fail.
\end{abstract}



\section{Introduction}\label{sec:p}

Let $(\Omega,\F,P)$ be a probability space with a filtration $\FF$ (an increasing sequence of sub-sigma-algebras of $\F$) and consider the problem
\begin{equation}\label{p}
\minimize\quad Ef(x(\omega),u(\omega),\omega)\quad\text{over $x\in\N$}
\end{equation}
where $f$ is an extended real-valued function, $\N$ is a space of $\FF$-adapted {\em decision strategies} and $u$ is a measurable function (exact definitions will be given below). The variable $u$ represents parameters or perturbations of a dynamic decision making problem where the objective is to minimize the expectation over {\em decision strategies} adapted to the information available to the decision maker over time. This paper derives dual expressions for the optimal value of \eqref{p} by incorporating some measure theoretic techniques from mathematical finance into the general conjugate duality framework of Rockafellar~\cite{roc74}.

Problem \eqref{p} covers many important optimization models in operations research and mathematical finance. Specific instances of stochastic optimization problems can often be put in the above format by appropriately specifying the integrand $f$. Allowing the integrand $f$ to take on the value $+\infty$, we can represent various pointwise (almost sure) constraints by infinite penalties. Some of the earliest examples can be found in Danzig~\cite{dan55} and Beale~\cite{bea55}. Problem~\eqref{p} provides a very general framework also for various optimization and pricing problems in mathematical finance. Certain classes of stochastic control problems can also put the above form; see \cite[Section~6]{rw78}. In some applications, the parameter $u$ is introduced into a given problem in order to derive information (such as optimality conditions or bounds on the optimal value) about it. This is the point of view taken e.g.\ in \cite{roc74}. In other applications, the parameter $u$ has a natural interpretation in the original formulation itself. Examples include financial applications where $u$ may represent the payouts of a financial instrument such as an option and one is trying to minimize the initial cost of a hedging portfolio. 

Convex duality has widespread applications in operations research, calculus of variations and mechanics. Besides in deriving optimality conditions, duality is used in numerical optimization and bounding techniques. The essence of convex duality is beautifully summarized by the conjugate duality framework of \cite{roc74} which subsumes more special duality frameworks such as Lagrangian (and in particular LP) and Fenchel duality; see also Ekeland and Temam~\cite{et76}. 
Several duality results, including optimality conditions for certain instances of \eqref{p} have been derived from the conjugate duality framework in Rockafellar and Wets~\cite{rw76,rw77,rw78,rw83}.

Convex duality has long been an integral part also of mathematical finance but there, duality results are often derived ad hoc instead of embedding a given problem in a general optimization framework. Attempts to derive financial duality results from known optimization frameworks are often hindered by two features. First, general duality frameworks are often formulated in locally convex topological vector spaces while in financial problems the decision strategies are usually chosen from a space that lacks an appropriate locally convex topology. Second, general duality results are often geared towards attainment of the dual optimum which requires conditions that often fail to hold in financial applications. The main contribution of this paper is to propose a general enough duality framework for \eqref{p} that covers several problems both in operations research as well as in mathematical finance. Our framework, to be rigorously specified in Section~\ref{sec:cd}, is an extension of the stochastic programming duality frameworks proposed in~\cite{rw76,rw78}. In our framework the parameters $u$ enter the model in a more general manner and we do not restrict the decision strategies $x$ to be bounded or integrable a priori. 

Allowing strategies to be general adapted processes has turned out be useful in deriving various duality results for financial models; see e.g.\ Schachermayer and Delbaen~\cite{ds6}, Kabanov and Safarian~\cite{ks9} and their references. This paper extends such techniques to a much more general class of models. We obtain dual representations for the optimal value of \eqref{p} but not necessarily the dual attainment as opposed to the strong duality results in \cite{rw76,rw77,rw78,rw83}. Consequently, we cannot claim the necessity of various optimality conditions involving dual variables. Nevertheless, the mere absence of duality gap is useful in many situations e.g.\ in mathematical finance where the ``constraint qualifications'' required for classical duality results often fail to hold. For example, various dual representations of hedging costs correspond to the absence of the duality gap while the dual optimum might not be attained. As an application, we extend certain results on superhedging and optimal consumption to a general market model with nonlinear illiquidity effects and convex portfolio constraints. This will be done by extending the elegant (currency) market model of Kabanov~\cite{kab99} where all assets are treated symmetrically. More traditional market models are then covered as special cases. The absence of duality gap is useful also in deriving certain simulation-based numerical techniques for bounding the optimum value of \eqref{p} as e.g.\ those proposed in Rogers~\cite{rog2} and Haugh and Kogan~\cite{hk4} in the case of optimal stopping problems. We extend such techniques for a more general class of problems.


The rest of this paper is organized as follows. Section~\ref{sec:cd} presents the general duality framework for problem \eqref{p} based on the conjugate duality framework of \cite{roc74}. Sections~\ref{sec:or} and \ref{sec:mf} give some well-known examples and extensions of duality frameworks from operations research and mathematical finance, respectively. Section~\ref{sec:cl} extends some classical closedness criteria from finite-dimensional spaces to the present infinite-dimensional stochastic setting.

\section{Conjugate duality}\label{sec:cd}

We study \eqref{p} in the {\em conjugate duality} framework of Rockafellar~\cite{roc74}. However, we deviate from \cite{roc74} in that the space $\N$ of decision variables need not be a locally convex topological vector space paired with another one. This precludes the completely symmetric duality in \cite{roc74} but in some situations it yields more regularity for the optimal value than what can  be obtained e.g.\ with integrable strategies. 

For given integers $n_t$, we set
\[
\N = \{(x_t)_{t=0}^T\,|\,x_t\in L^0(\Omega,\F_t,P;\reals^{n_t})\},
\]
where $L^0(\Omega,\F_t,P;\reals^{n_t})$ denotes the space of equivalence classes of $\F_t$-measurable $\reals^{n_t}$-valued functions that coincide $P$-almost surely. Each $x_t$ is interpreted as a decision that is made after observing all available information at time $t$. In applications, the filtration $\FF$ is often generated by a finite-dimensional stochastic process whose values are observed at discrete points in time. If $\F_0$ is the trivial sigma algebra $\{\emptyset,\Omega\}$ then the first component $x_0$ is deterministic, 

The function $f$ is assumed to be an {\em extended real-valued convex normal integrand} on $\reals^n\times\reals^m\times\Omega$ where $n=n_0+\ldots+n_T$ and $m$ is a given integer. This means that the set-valued mapping $\omega\mapsto\epi f(\cdot,\cdot,\omega)$ is $\F$-measurable and it has closed and convex values (so $(x,u)\mapsto f(x,u,\omega)$ is convex and lower semicontinuous for every $\omega$); see e.g.\ \cite[Chapter~14]{rw98}. This implies that $f$ is $\B(\reals^n\times\reals^m)\otimes\F$-measurable and that the function $(x,u)\mapsto f(x,u,\omega)$ is lower semicontinuous and convex for every $\omega$. 
It follows that $\omega\mapsto f(x(\omega),u(\omega),\omega)$ is $\F$-measurable for every $x\in L^0(\Omega,\F,P;\reals^n)$ and $u\in L^0(\Omega,\F,P;\reals^m)$. Throughout this paper, the expectation of an extended real-valued measurable function is defined as $+\infty$ unless the positive part is integrable. The {\em integral functional}
\[
I_f(x,u) := Ef(x(\omega),u(\omega),\omega)
\]
in the objective of \eqref{p} is then well-defined extended real-valued convex function on $L^0(\Omega,\F,P;\reals^n)\times L^0(\Omega,\F,P;\reals^m)$. Normal integrands possess many useful properties and they arise quite naturally in many optimization problems in practice. Examples will be given in the following sections. We refer the reader to \cite{roc76} or \cite[Chapter 14]{rw98} for general treatment of normal integrands on $\reals^d\times\Omega$ for finite $d$.

For each $u\in L^0(\Omega,\F,P;\reals^m)$, the optimal value of \eqref{p} is given by the {\em value function}
\[
\varphi(u) := \inf_{x\in\N}I_f(x,u).
\]
By \cite[Theorem~1]{roc74}, $\varphi$ is convex. We will derive dual expressions for $\varphi$ on the space $L^p := L^p(\Omega,\F,P;\reals^m)$ using the conjugate duality framework of Rockafellar~\cite{roc74}. To this end, we pair $L^p$ with $L^q$, where $q\in[1,\infty]$ is such that $1/p+1/q=1$. The bilinear form
\[
\langle u,y\rangle = E[u(\omega)\cdot y(\omega)]
\]
puts $L^p$ and $L^q$ in separating duality. The weakest and the strongest locally convex topologies on $L^p$ compatible with the pairing will be denoted by $\sigma(L^p,L^q)$ and $\tau(L^p,L^q)$, respectively (similarly for $L^q$). By the classical separation argument, a convex function is lower semicontinuous with respect to $\sigma(L^p,L^q)$ if it is merely lower semicontinuous with respect to $\tau(L^p,L^q)$.

\begin{remark}
For $p\in[1,\infty)$, $\tau(L^p,L^q)$ is the norm topology and $\sigma(L^p,L^q)$ is the weak-topology that $L^p$ has as a Banach space with the usual $L^p$-norm. For $p=\infty$, $\sigma(L^p,L^q)$ is the weak*-topology that $L^p$ has as the Banach dual of $L^q$ while $\tau(L^p,L^q)$ is, in general, weaker than the norm topology. It follows from the Mackey-Arens and Dunford-Pettis theorems, that a sequence in $L^\infty$ converges with respect to $\tau(L^\infty,L^1)$ if and only if it norm-bounded and converges in measure; see Grothendieck~\cite[Part~4]{gro73} for the case of locally compact measure spaces. In mathematical finance, a convex function on $L^\infty$ is sometimes said to have the ``Fatou property'' if it is sequentially lower-semicontinuous with respect to $\tau(L^\infty,L^1)$.
\end{remark}

\begin{remark}
Instead of $L^p$ and $L^q$, we could take an arbitrary pair of spaces of measurable $\reals^m$-valued functions which are in separating duality under the bilinear form $\langle u,y\rangle = E[u(\omega)\cdot y(\omega)]$. Examples include Orlicz spaces which have recently been used in a financial context by Biagini and Frittelli~\cite{bf8}.
\end{remark}

The {\em conjugate} of a function $\varphi$ on $L^p$ is the convex function on $L^q$ defined by
\[
\varphi^*(y) = \sup_{u\in L^p}\{\langle u,y\rangle - \varphi(u)\}.
\]
The conjugate of a function on $L^q$ is defined similarly. It is a fundamental result in convex duality that $\varphi^{**}=\cl\varphi$ where
\[
\cl\varphi = 
\begin{cases}
\lsc\varphi & \text{if $(\lsc\varphi)(u) > -\infty\ \forall u\in L^p$},\\
-\infty & \text{otherwise}
\end{cases}
\]
is the {\em closure} of $\varphi$; see e.g.\ \cite[Theorem~5]{roc74}. Here $\lsc\varphi$ denotes the {\em lower semicontinuous hull} of $\varphi$. If $\lsc\varphi$ has a finite value at some point then $\lsc\varphi$ is proper and $\lsc\varphi = \cl\varphi$; see \cite[Theorem~4]{roc74}.

The {\em Lagrangian} associated with \eqref{p} is the extended real-valued function on $\N\times L^q$ defined by
\[
L(x,y) = \inf_{u\in L^p}\{I_f(x,u)-\langle u,y\rangle\}.
\]
The Lagrangian is convex in $x$ and concave in $y$. The {\em dual objective} is the extended real-valued function on $L^q$ defined by 
\[
g(y) = \inf_{x\in\N}L(x,y).
\]
Since $g$ is the pointwise infimum of concave functions, it is concave. The basic duality result \cite[Theorem~7]{roc74} says, in particular, that
\[
g = -\varphi^*.
\]
This follows directly from the above definitions and does not rely on topological properties of $\N$. The biconjugate theorem then gives the dual representation
\begin{equation}\label{bc}
(\cl\varphi)(u) = \sup\{\langle u,y\rangle + g(y)\}.
\end{equation}

In many applications, the parameter $u$ has practical significance, and the dual representation \eqref{bc} may yield valuable information about the function $\varphi$. On the other hand, in some situations, one is faced with a fixed optimization problem and the parameter $u$ is introduced in order to derive information about the original problem. This is the perspective taken in \cite{roc74}, where the minimization problem
\begin{equation}\label{primal}
\minimize\quad I_f(x,0)\quad\text{over}\quad x\in\N
\end{equation}
would be called the {\em primal problem} and 
\begin{equation}\label{dual}
\maximize\quad g(y)\quad\text{over}\quad y\in L^q
\end{equation}
the {\em dual problem}. By \eqref{bc}, the optimum values of \eqref{primal} and \eqref{dual} are equal exactly when $(\cl\varphi)(0)=\varphi(0)$. An important topic which is studied in \cite{roc74} but not in the present paper is derivatives of the value function $\varphi$ and the associated optimality conditions. In this paper, we concentrate on the more general property of lower semicontinuity of $\varphi$; see Section~\ref{sec:cl}. The lower semicontinuity already yields many interesting results in operations research and mathematical finance. Moreover, lower semicontinuity is useful for proving the continuity of $\varphi$ for $p<\infty$ since a lower semicontinuous convex function on a barreled space is continuous throughout the interior of its domain; see e.g.\ \cite[Corollary~8B]{roc74}. 

\begin{remark}
As long as the integral functional $(x,u)\mapsto I_f(x,u)$ is closed in $u$ (which holds under quite general conditions given e.g.\ in Rockafellar~\cite{roc68}), the biconjugate theorem gives
\[
I_f(x,u) = \sup_{y\in L^q}\{L(x,y)+\langle u,y\rangle\}
\]
and, in particular, $I_f(x,0)=\sup_yL(x,y)$ so that $\varphi(0)=\inf_{x\in\N}\sup_{y\in L^q} L(x,y)$. On the other hand, \eqref{bc} gives $(\cl\varphi)(0)=\sup_{y\in L^q}\inf_{x\in\N}L(x,y)$ so that the condition $(\cl\varphi)(0)=\varphi(0)$ can be expressed as
\[
\inf_{x\in\N}\sup_{y\in L^q} L(x,y)=\sup_{y\in L^q}\inf_{x\in\N}L(x,y).
\]
 In other words, the function $L$ has a saddle-value iff $\varphi$ is closed at the origin. Along with the general duality theory for convex minimization, the conjugate duality framework of \cite{roc74} addresses general convex-concave minimax problems.
\end{remark}

The following interchange rule will be useful in deriving more explicit expressions for the dual objective $g$. It is a special case of \cite[Theorem~14.60]{rw98} and it uses the fact that for an $\F$-measurable normal integrand $h$, the function $\omega\mapsto\inf_u h(u,\omega)$ is $\F$-measurable; see \cite[Theorem~14.37]{rw98}. 

\begin{theorem}[Interchange rule]\label{int}
Given an $\F$-measurable normal integrand $h$ on $\reals^k\times\Omega$, we have
\[
\inf_{u\in L^p}Eh(u(\omega),\omega) = E\inf_{u\in\reals^k}h(u,\omega)
\]
as long as the left side is less than $+\infty$.
\end{theorem}

Theorem~\ref{int} yields a simple proof of Jensen's inequality. Throughout this paper, the {\em conditional expectation} of a random variable $x$ with respect to $\F_t$ will be denoted by $E_tx$; see e.g.\ Shiryaev~\cite[II.7]{shi96}.

\begin{corollary}[Jensen's inequality]
Let $h$ is an $\F_t$-measurable convex normal integrand on $\reals^k\times\Omega$ such that $Eh^*(v(\omega),\omega)<\infty$ for some $v\in L^q(\Omega,\F_t,P;\reals^k)$. Then
\[
Eh((E_tx)(\omega),\omega)\le Eh(x(\omega),\omega)
\]
for every $x\in L^p(\Omega,\F,P;\reals^k)$.
\end{corollary}

\begin{proof}
Applying Theorem~\ref{int} twice, we get 
\begin{align*}
I_h(E_tx) &= E\sup_{v}\{v\cdot(E_tx)(\omega)-h^*(v,\omega))\}\\
&=\sup_{v\in L^q(\F_t)}E\{v(\omega)\cdot(E_tx)(\omega)-h^*(v(\omega),\omega)\}\\
&=\sup_{v\in L^q(\F_t)}E\{v(\omega)\cdot x(\omega)-h^*(v(\omega),\omega)\}\\
&\le\sup_{v\in L^q(\F)}E\{v(\omega)\cdot x(\omega)-h^*(v(\omega),\omega)\}\\
&= E\sup_v\{v\cdot x(\omega)-h^*(v,\omega)\}\\
&=Eh(x(\omega),\omega),
\end{align*}
where the third equality comes from the law of iterated expectations; see e.g.~\cite[Section~II.7]{shi96}.
\end{proof}

Going back to \eqref{p}, we define
\[
l(x,y,\omega) = \inf_{u\in\reals^m}\{f(x,u,\omega) - u\cdot y\}.
\]
This is an extended real-valued function on $\reals^n\times\reals^m\times\Omega$, convex in $x$ and concave in $y$. Various dual expressions in stochastic optimization and in mathematical finance can be derived from the following result which expresses the dual objective in terms of $l$. In many situations, the expression can be written concretely in terms of problem data; see Sections~\ref{sec:or} and \ref{sec:mf}. Given an $r\in[1,\infty]$, we let
\[
\N^r:=\N\cap L^r(\Omega,P,\F;\reals^n).
\]

\begin{theorem}\label{thm}
The function $\omega\mapsto l(x(\omega),y(\omega),\omega)$ is measurable for any $x\in\N$ and $y\in L^q$ so the integral functional $I_l(x,y)=El(x(\omega),y(\omega),\omega)$ is well-defined on $\N\times L^q$. As long as $I_f\not\equiv+\infty$, we have
\[
g(y) = \inf_{x\in\N} I_l(x,y)
\]
If, in addition, $l$ is of the form\footnote{Throughout this paper, we define $\infty-\infty=+\infty$.}
\[
l(x,y,\omega) = \sum_{t=0}^Tl_t(x_t,y,\omega)
\]
for some $\B(\reals^{n_t})\otimes\B(\reals^m)\otimes\F$-measurable extended real-valued functions $l_t$ on $\reals^{n_t}\times\reals^m\times\Omega$ then
\[
g(y) = \inf_{x\in\N^r}I_l(x,y)
\]
as long as the right side is less than $+\infty$.
\end{theorem}

\begin{proof}
We have $l(x(\omega),y(\omega),\omega)=-h^*(y(\omega),\omega)$, where $h(u,\omega):=f(x(\omega),u,\omega)$. To prove the measurability it suffices to show that $h^*$ is a normal integrand on $\reals^m\times\Omega$. This follows from Proposition~14.45(c) and Theorem~14.50 of \cite{rw98}.

If $I_f\not\equiv+\infty$, then there exists an $x\in\N$ such that $L(x,y)<\infty$ for every $y\in L^q$. We can thus assume that $L(x,y)<\infty$ in the expression for $g$ in which case
\[
L(x,y) = E\inf_{u\in\reals^m}\{f(x(\omega),u,\omega)-u\cdot y(\omega)\} = I_l(x,y),
\]
by Theorem~\ref{int}. Here we apply the interchange rule to the function $(u,\omega)\mapsto f(x(\omega),u,\omega)$ which is a normal integrand, by \cite[Proposition~14.45(c)]{rw98}.

Fix a $y\in L^q$ and let $x\in\N^r$ be such that $I_l(x,y)<\infty$. Let $\eps>0$ be arbitrary and let $x'\in\N$ be such that $El(x'(\omega),y(\omega),\omega)\le g(y)+\eps$. Defining $x_t^\nu = x_t'\chi_{A_t^\nu} + x_t\chi_{\bar\Omega\setminus A_t^\nu}$, where $A_t^\nu=\{\omega\,|\,|x'_t(\omega)|\le\nu\}$, we have that the strategy $x^\nu=(x_t^\nu)_{t=0}^T$ is in $\N^r$ and that $x_t^\nu\to x_t'$ almost surely  for every $t=0,\ldots,T$ as $\nu\upto\infty$. Since the functions $\omega\mapsto l_t(x_t^\nu(\omega),y(\omega),\omega)$ are dominated by the integrable function 
\[
\omega\mapsto\max\{l_t(x'_t(\omega),y(\omega),\omega),l_t(x_t(\omega),y(\omega),\omega),0\},
\]
Fatou's lemma (applied in the product measure space $\Omega\times\{0,\ldots,T\}$ obtained by equipping $\{0,\ldots,T\}$ with the counting measure) gives
\begin{align*}
\limsup E\sum_{t=0}^Tl_t(x_t^\nu(\omega),y(\omega),\omega) &\le E\sum_{t=0}^T\limsup l_t(x_t^\nu(\omega),y(\omega),\omega)\\
&= E\sum_{t=0}^Tl_t(x'_t(\omega),y(\omega),\omega)\\
&\le g(y) +\eps.
\end{align*}
Since $\eps>0$ was arbitrary and $x^\nu\in\N^r$, the claim follows.
\end{proof}

The main content of the first part of Theorem~\ref{thm} is that the infimum in the definition of the Lagrangian can be reduced to scenariowise minimization. This can sometimes be done even analytically. The last part of the above result shows that, while integrability of $x$ may be restrictive in the original problem, it may be harmless in the expression for the dual objective $g$. A simple example will be given at the end of Example~\ref{ex:ie} below. In some applications, the integrability can be used to derive more convenient expressions for $g$.




\section{Examples from operations research}\label{sec:or}

This section reviews some well-known duality frameworks from operations research and shows how they can be derived from the abstract framework above. Many of the examples are from Rockafellar and Wets~\cite{rw76,rw78} where they were formulated for bounded strategies. We will also point out some  connections with more recent developments in finance and stochastics. A recent account of techniques and models of stochastic programming can be found in Shapiro, Dentcheva and Ruszczyński~\cite{sdr9}.

The best known duality frameworks involve functional constraints and Lagrange multipliers. The most classical example is linear programming duality. These frameworks are deterministic special cases of the following stochastic programming framework from \cite{rw78}, where sufficient conditions were given for the attainment of the dual optimum.

\begin{example}[Inequality constraints]\label{ex:ie}
Let
\[
f(x,u,\omega) =
\begin{cases}
f_0(x,\omega) & \text{if $f_j(x,\omega)+u_j\le 0$ for $j=1,\ldots,m$},\\
+\infty & \text{otherwise},
\end{cases}
\]
where $f_j$ are convex normal integrands. To verify that $f$ is a normal integrand, we write it as $f = f_0 + \sum_{j=1}^m\delta_{C_j}$, where 
\[
\delta_{C_j}(x,u,\omega)=
\begin{cases}
0 & \text{if $(x,u)\in C_j(\omega)$},\\
+\infty & \text{otherwise}
\end{cases}
\]
and $C_j(\omega)=\{(x,u)\,|\, f_j(x,\omega)+u_j\le 0\}$. By \cite[Proposition~14.33]{rw98}, the sets $C_j$ are measurable so the functions $\delta_{C_j}$ are normal integrands by \cite[Example~14.32]{rw98} and then $f$ is a normal integrand by \cite[Proposition~14.44(c)]{rw98}. The integral functional $I_f$ is thus well-defined and equals
\[
I_f(x,u) = 
\begin{cases}
Ef_0(x(\omega),\omega) & \text{if $f_j(x(\omega),\omega)+u(\omega)\le 0\ P\text{-a.s.}\ j=1,\ldots,m$},\\
+\infty & \text{otherwise}.
\end{cases}
\]
The primal problem \eqref{primal} can be written as
\begin{alignat*}{2}
&\minimize\quad& E&f_0(x(\omega),\omega)\quad\text{{\rm over} $x\in\N$}\\
&\st\quad& &f_j(x(\omega),\omega)\le 0\ P\text{-a.s.},\ j=1,\ldots,m.
\end{alignat*}
This is the classical formulation of a nonlinear stochastic optimization problem. It is a stochastic extension of classical mathematical programming models such as linear programming.

The Lagrangian integrand becomes
\begin{align*}
l(x,y,\omega) &= \inf_{u\in\reals^m}\{f(x,u,\omega)-u\cdot y\}\\
&=
\begin{cases}
+\infty & \text{if $f_j(x,\omega)=\infty$ for some $j$},\\
f_0(x,\omega) + y\cdot F(x,\omega) & \text{if $f_j(x,\omega)<\infty$ and $y\ge 0$},\\
-\infty & \text{otherwise},
\end{cases}
\end{align*}
where $F(x,\omega)=(f_1(x,\omega),\ldots,f_m(x,\omega))$. The expression $g(y)=\inf_{x\in\N}I_l(x,y)$ holds under the general condition of Theorem~\ref{thm}, but to get more explicit expressions for the dual objective $g$ one needs more structure on $f$; see the examples below. 

To illustrate how the choice of the strategy space may affect the lower semicontinuity of $\varphi$, consider the case $n=m=1$, $f_0=0$ and 
\[
f_1(x,u,\omega)= a(\omega)x + u,
\]
for some strictly positive $a$ such that $1/a\notin L^1$. We get $\varphi(u)=0$ for every $u\in L^p$ but there is no $x\in\N^1$ which satisfies the pointwise constraint when $\essinf u>0$. However,
\[
l(x,y,\omega) = 
\begin{cases}
ya(\omega)x & \text{if $y\ge 0$},\\
-\infty & \text{otherwise},
\end{cases}
\]
so, by the second part of Theorem~\ref{thm}, the strategies can be taken even bounded when calculating $g$.
\end{example}

It was observed in \cite[Section~3A]{rw78} that the dual objective in Example~\ref{ex:ie} can be written in a more concrete form when the functions $f_j$ have a time-separable form.

\begin{example}\label{ex:ies}
Consider Example~\ref{ex:ie} in the case
\[
f_j(x,\omega)=\sum_{t=0}^Tf_{j,t}(x_t,\omega),
\]
where each $f_{j,t}$ is an $\F_t$-measurable normal integrand. Defining $F_t(x_t,\omega)=(f_{1,t}(x_t,\omega),\ldots,f_{m,t}(x_t,\omega))$ and using the convention $\infty-\infty=+\infty$, we can write
\[
l(x,y,\omega) = \sum_{t=0}^Tl_t(x_t,y,\omega),
\]
where
\[
l_t(x_t,y,\omega) = 
\begin{cases}
+\infty & \text{if $f_{j,t}(x_t,\omega)=\infty$ for some $j$},\\
f_{0,t}(x_t,\omega) + y\cdot F_t(x_t,\omega) & \text{if $f_{j,t}(x_t,\omega)<\infty$ and $y\ge 0$},\\
-\infty & \text{otherwise}.
\end{cases}
\]

Assume now that $F_t(x_t,\cdot)\in L^p$ for every $t$ and $x_t\in\reals^{n_t}$ and that there is a $v\in\N^p$ and a $p$-integrable random variable $w$ such that $f_{j,t}(x,\omega)\ge v_t(\omega)\cdot x - w(\omega)$. It follows that $F(x(\cdot),\cdot)\in L^p$ for every $x\in\N^\infty$; see e.g.\ \cite[Theorem~3K]{roc76}\footnote{If $\|x\|_{L^\infty}\le r$, there is a finite set of points $x^i\in\reals^J$ $i=1,\ldots,n$ whose convex combination contains the ball $r\uball$. By convexity, $f_{j,t}(z(\omega),\omega)\le\sup_{i=1,\ldots,n}f_{j,t}(x^i,\omega)$, where the right hand side is $p$-integrable by assumption. Combined with the lower bound, we then have $F_t(x(\cdot),\cdot)\in L^p$ as claimed.}. If there is an $x\in\N^\infty$ such that $\omega\mapsto f_{0,t}(x_t(\omega),\omega)$ are integrable then, by the second part of Theorem~\ref{thm}, 
\[
g(y) = \inf_{x\in\N^\infty}E\sum_{t=0}^Tl_t(x_t(\omega),y(\omega),\omega).
\]
Using the properties of conditional expectation (see e.g.\ \cite[Section~II.7]{shi96}), we get
\begin{align*}
g(y) &= \inf_{x\in\N^\infty}E\sum_{t=0}^TE_tl_t(x_t(\omega),y(\omega),\omega)\\
&= \inf_{x\in\N^\infty}E\sum_{t=0}^Tl_t(x_t(\omega),(E_ty)(\omega),\omega).
\end{align*}
Applying Theorem~\ref{int} for $t=0,\ldots,T$, we can express the dual objective as
\begin{align*}
g(y) &= E\sum_{t=0}^Tg_t((E_ty)(\omega),\omega),
\end{align*}
where
\[
g_t(y,\omega)=\inf_{x_t\in\reals^{n_t}}l_t(x_t,y,\omega).
\]
The dual problem can thus be written as
\[
\maximize_{y\in\M^p}\quad \sum_{t=0}^Tg_t(y_t(\omega),\omega),
\]
where $\M^q$ is the set of $\reals^m$-valued $q$-integrable martingales.
\end{example}

In the linear case, considered already in Danzig~\cite{dan55}, the dual problem in Example~\ref{ex:ies} can be written as another linear optimization problem. 

\begin{example}[Linear programming]\label{ex:lp}
Consider Example~\ref{ex:ies} in the case where 
\[
f_{0,t}(x_t,\omega)=
\begin{cases}
a_{0,t}(\omega)\cdot x_t & \text{if $x\in\reals^{n_t}_+$,}\\
+\infty & \text{otherwise}
\end{cases}
\]
and $f_{j,t}(x_t,\omega)=a_{j,t}(\omega)\cdot x_t + b_{j,t}(\omega)$ for $\F_t$-measurable $p$-integrable $n_t$-dimensional vectors $a_{j,t}$ and $\F_t$-measurable integrable scalars $b_{j,t}$. The primal problem can then be written as 
\begin{alignat*}{2}
&\minimize\quad& E&\sum_{t=0}^Ta_{0,t}(\omega)\cdot x_t(\omega)\quad\text{{\rm over} $x\in\N_+$}\\
&\st\quad& &\sum_{t=0}^T[A_t(\omega)x_t(\omega)+b_t(\omega)]\le 0\quad P\text{-a.s.},
\end{alignat*}
where $A_t(\omega)$ is the matrix with rows $a_{j,t}(\omega)$ and $b_t(\omega)=(b_{j,t}(\omega))_{j=1}^m$. We get
\begin{align*}
l_t(x_t,y,\omega)
&=
\begin{cases}
+\infty & \text{if $x_t\not\ge 0$},\\
a_{0,t}(\omega)\cdot x_t + y\cdot[A_t(\omega) x_t + b_t(\omega)]& \text{if $x_t\ge 0$, $y\ge 0$},\\
-\infty & \text{otherwise}
\end{cases}\\
&=
\begin{cases}
+\infty & \text{if $x_t\not\ge 0$},\\
[A^*_t(\omega) y+a_{0,t}(\omega)]\cdot x_t + y\cdot b_t(\omega)& \text{if $x_t\ge 0$, $y\ge 0$},\\
-\infty & \text{otherwise},
\end{cases}
\end{align*}
where $A_t^*(\omega)$ is the transpose of $A_t(\omega)$. It follows that
\[
g_t(y,\omega) = 
\begin{cases}
y\cdot b_t(\omega) & \text{if $y\ge 0$ and $A^*_t(\omega) y+a_{0,t}(\omega)\ge 0$},\\
-\infty & \text{otherwise}
\end{cases}
\]
and 
the dual problem can be written as
\begin{alignat*}{2}
&\minimize\quad& E&\sum_{t=0}^Tb_t(\omega)\cdot y_t(\omega)\quad\text{{\rm over} $y\in\M^q_+$}\\
&\st\quad& & A_t^*(\omega)y_t(\omega)+a_{0,t}(\omega)\ge 0\quad P\text{-a.s.}\ t=0,\ldots,T,
\end{alignat*}
where $\M^q_+$ is the set of nonnegative $q$-integrable martingales. When $T=0$, we recover the classical linear programming duality framework.
\end{example}

The famous problem of optimal stopping is a one-dimensional special case of Example~\ref{ex:lp}.

\begin{example}[Optimal stopping]\label{ex:os}
The optimal stopping problem with an integrable nonnegative scalar process $Z$ can be formulated as
\[
\maximize_{x\in\N_+}\quad E\sum_{t=0}^T x_tZ_t\quad\st\quad\sum_{t=0}^T x_t\le 1,\ x_t\in\{0,1\}\ P\text{-a.s.}
\]
The feasible strategies $x$ are related to stopping times through $\tau(\omega) = \inf\{t\,|\, x_t(\omega)=1\}$. The optimal value is not affected if we relax the constraint $x_t\in\{0,1\}$  (see below). The relaxed problem fits the framework of Example~\ref{ex:lp} with $n_t=m=1$, $p=\infty$, $a_{0,t}(\omega)=-Z_t(\omega)$, $a_{1,t}(\omega)=1$ and $b_{1,t}(\omega)=-1/(T+1)$. The dual problem becomes 
\[
\minimize_{y\in\M^1}\quad Ey_0\quad\st\quad y\ge Z\quad P\text{-a.s.}.
\]

To justify the convex relaxation, we first note that the feasible set of the relaxed problem is contained in the space $\N^\infty$ of bounded strategies. Since $Z\in\N^1$ by assumption, it suffices (by the Krein-Millman theorem) to show that the feasible set of the relaxed problem equals the $\sigma(\N^\infty,\N^1)$-closed convex hull of the feasible set of the original problem. Let $x$ be feasible in the relaxed problem. For $\nu=1,2,\ldots$, define the stopping times
\[
\tau^{\nu,i}(\omega) = \inf\{s\,|\,X_s(\omega)\ge i/\nu\}\quad i=1\ldots,\nu,
\]
where $X_s(\omega)=\sum_{t=0}^sx_t(\omega)$. The strategies 
\[
x^{\nu,i}_t(\omega) = 
\begin{cases}
1 & \text{if $\tau^{\nu,i}(\omega)=t$},\\
0 & \text{otherwise}
\end{cases}
\]
are feasible in the original problem. It suffices to show that the convex combinations 
\[
x^\nu(\omega) = \sum_{i=1}^\nu\frac{1}{\nu}x^{\nu,i}(\omega)
\]
converge to $x$ in the weak topology. By construction, 
\[
X^\nu_s(\omega) := \sum_{t=0}^sx^\nu_t(\omega) = \sup\{i\,|\,i/\nu\le X_s(\omega)\}\in[X_t(\omega)-1/\nu,X_t(\omega)],
\]
so that $X^\nu_t\to X_t$ and thus $x^\nu_t\to x_t$ almost surely. Since $x^\nu$ and $x$ are all contained in the unit ball of $\N^\infty$, we have
\[
E\sum_{t=0}^T x^\nu_tv_t\to E\sum_{t=0}^T x_tv_t\quad\forall v\in\N^1,
\]
by the dominated convergence theorem.
\end{example}

\begin{remark}\label{rem:sim}
The above duality frameworks suggest computational techniques for estimating the optimal value of the primal problem. The dual objective in Example~\ref{ex:ies} is dominated for every $y\in\M^q_+$ by
\begin{align}
\tilde g(y)&:=
E\inf_{x\in\reals^n}\left\{\left.\sum_{t=0}^T[f_t(x_t,\omega) + y_t(\omega)\cdot F_t(x_t,\omega)] \,\right|\, \sum_{t=0}^TF_t(x_t,\omega)\le 0\right\}.\label{intermediate}
\end{align}
If $x'\in\N$ is feasible in the primal problem, we get for every $y\in\M^q_+$
\begin{align*}
\tilde g(y) &\le E\sum_{t=0}^T[f_t(x'_t(\omega),\omega)+y_t(\omega)\cdot F_t(x'_t(\omega),\omega)] \\
&=E\left\{\sum_{t=0}^Tf_t(x'_t(\omega),\omega) + y_T(\omega)\cdot\sum_{t=0}^T F_t(x'_t(\omega),\omega)\right\}\\
&\le E\sum_{t=0}^Tf_t(x'_t(\omega),\omega).
\end{align*}
Minimizing over all feasible strategies $x'\in\N$ shows that \eqref{intermediate} lies between $g(y)$ and the optimum primal value $\varphi(0)$. When $\varphi$ is closed, we thus get that $\varphi(0)=\sup_{y\in\M^q_+}\tilde g(y)$. The problem of finding the infimum in \eqref{intermediate} can be seen as a deterministic version of the primal problem augmented by a penalty term in the objective. 

In the case of Example~\ref{ex:os}, \eqref{intermediate} can be written for every $y\in\M^q_+$ as
\begin{align*}
\tilde g(y) &= E\inf_{x\in\reals^n}\left\{\left.\sum_{t=0}^T[-Z_tx_t+y_t(x_t-1/(T+1))]\,\right|\,\sum_{t=0}^Tx_t\le 1\right\}\\
&= E\inf_{x\in\reals^n}\left\{\left.\sum_{t=0}^T[(y_t-Z_t)x_t-y_0]\,\right|\,\sum_{t=0}^Tx_t\le 1\right\}\\
&= E\min_{t=0,\ldots,T}(y_t-y_0-Z_t).
\end{align*}
This is the dual representation for optimal stopping obtained by Davis and Karatzas~\cite{dk94}. This was used by Rogers~\cite{rog2} (see also Haugh and Kogan~\cite{hk4}) in a simulation based technique for computing upper bounds for the value of American options in complete market models. The technique is readily extended to the more general problem class of Example~\ref{ex:ies}. The technique can be further extended using the following.
\end{remark}

The cost of the nonanticipativity constraint on the strategies has been studied in a number of papers; see e.g.\ Rockafellar and Wets~\cite{rw76} for a general discrete finite time framework as well as Wets~\cite{wet75}, Back and Pliska~\cite{bp87}, Davis~\cite{dav92} and Davis and Burnstein~\cite{db92} on continuous-time models. The cost can be described in terms of dual variables representing the value of information. The following derives a dual representation in the framework of Section~\ref{sec:cd}.

\begin{example}[Shadow price of information]\label{ex:ic}
Let $h$ be a convex normal integrand and consider the problem
\begin{equation}\label{sph}
\minimize_{x\in\N}\quad I_h(x).
\end{equation}
This can be seen as the primal problem associated with the normal integrand
\[
f(x,u,\omega) = h(x+u,\omega).
\]
The value function $\varphi(u)$ corresponds to adding a general $\F_T$-measurable vector $u_t$ to each $x_t$ in \eqref{sph}. We get
\begin{align*}
l(x,y,\omega) &= \inf_{u\in\reals^m}\{h(x+u,\omega)-u\cdot y\}\\
&= \inf_{w\in\reals^m}\{h(w,\omega)-\sum_{t=0}^T(w_t-x_t)\cdot y_t\}\\
&= \sum_{t=0}^Tx_t\cdot y_t -\sup_{w\in\reals^m}\{\sum_{t=0}^Tw_t\cdot y_t-h(w,\omega)\}\\
&= \sum_{t=0}^Tx_t\cdot y_t - h^*(y,\omega).
\end{align*}
As long as there is a $y\in L^q$ such that $I_{h^*}(y)<\infty$, this satisfies the conditions of Theorem~\ref{thm} with $r=p$ so that
\begin{align*}
g(y) &= \inf_{x\in\N^p}E\{\sum_{t=0}^Tx_t(\omega)\cdot y_t(\omega) - h^*(y(\omega),\omega)\}\\
&= 
\begin{cases}
-Eh^*(y(\omega),\omega) & \text{if $y\perp\N^p$},\\
-\infty & \text{otherwise}.
\end{cases}
\end{align*}

By Theorem~\ref{thm},
\begin{align*}
(\cl\varphi)(0) &= \sup_{y\perp\N^p}-Eh^*(y(\omega),\omega)\\
&= \sup_{y\perp\N^p}E\inf_{x\in\reals^n}\{h(x,\omega)-\sum_{t=0}^Tx_t\cdot y_t(\omega)\}.
\end{align*}
The infimum in the last expression differs from the original problem in that the information constraints have been replaced by a linear term. This can be used to compute lower bounds for the optimal value using simulation much like in Rogers~\cite{rog2} and Haugh and Kogan~\cite{hk4} in the case of optimal stopping problems; see Remark~\ref{rem:sim}. 
Rockafellar and Wets~\cite{rw76} gave sufficient conditions for the existence of a $y\in\M^1$ such that $\varphi(0)=g(y)$ in the case of bounded strategies; see also Back and Pliska~\cite{bp87} for a continuous-time framework with a special class of objective functions. 
\end{example}

The following problem format is adapted from Rockafellar and Wets~\cite{rw83}. It has its roots in calculus of variations and optimal control; see Rockafellar~\cite{roc71}.

\begin{example}[Problems of Bolza type]\label{ex:bolza}
Let $n_t=d$ and consider the problem
\begin{equation}\label{bolza}
\minimize_{x\in\N}\quad E\sum_{t=0}^TL_t(x_t(\omega),\Delta x_t(\omega),\omega),
\end{equation}
where $\Delta x_t:=x_t-x_{t-1}$, $x_{-1}:=0$ and each $L_t$ is an $\F_t$-measurable normal integrand on $\reals^d\times\reals^d\times\Omega$. This fits our general framework with
\[
f(x,u,\omega) = \sum_{t=0}^TL_t(x_t,\Delta x_t+u_t,\omega),
\]
where $x_{-1}:=0$ and $u=(u_0,\ldots,u_T)$ with $u_t\in\reals^d$. Indeed, \eqref{bolza} is \eqref{p} with $u=0$. We get
\begin{align*}
l(x,y,\omega) &= \inf_{u\in\reals^m}\sum_{t=0}^T[L_t(x_t,\Delta x_t+u_t,\omega)-u_t\cdot y_t]\\
&= \inf_{v\in\reals^m}\sum_{t=0}^T[L_t(x_t,v_t,\omega)-(v_t-\Delta x_t)\cdot y_t]\\
&= \sum_{t=0}^T[\Delta x_t\cdot y_t-H_t(x_t,y_t,\omega)]\\
&= \sum_{t=0}^T[-x_t\cdot\Delta y_{t+1}-H_t(x_t,y_t,\omega)],
\end{align*}
where $y_{T+1}:=0$ and $H_t$ is the {\em Hamiltonian} defined by 
\[
H_t(x_t,y_t,\omega) = \sup_{v_t\in\reals^d}\{v_t\cdot y_t-L_t(x_t,v_t,\omega)\}.
\]
Thus
\begin{align*}
g(y) &= \inf_{x\in\N}E\sum_{t=0}^T\left[\Delta x_t\cdot y_t - H_t(x_t,y_t)\right].
\end{align*}
By Jensen's inequality, $g(y)\le g(\pi y)$ where $\pi$ denotes the projection $(y_t)_{t=0}^T\mapsto(E_ty_t)_{t=0}^T$. Consequently, when maximizing $g$, we do not loose anything if we restrict $y$ to the space $\N^q$ of adapted $q$-integrable processes. Moreover, if $u\in\N^p$ we have $\langle u,y\rangle=\langle u,\pi y\rangle$, so that
\begin{align*}
(\cl\varphi)(u) &= \sup_{y\in L^q}\{\langle u,y\rangle + g(y)\} = \sup_{y\in\N^q}\{\langle u,y\rangle + g(y)\}.
\end{align*}

Assume now that there is an $x\in\N^p$ such that $EH_t(x_t,y)<\infty$ and that $(x_t,\omega)\mapsto -H_t(x_t,y(\omega),\omega)$ are $\F_t$-measurable normal integrands for every $y\in\N^q$. We then get from Theorem~\ref{thm}, the law of iterated expectations (see e.g.~\cite[Section~II.7]{shi96}) and Theorem~\ref{int} that for every $y\in\N^q$
\begin{align*}
g(y) &= \inf_{x\in\N^p}E\sum_{t=0}^T\left[-x_t\cdot\Delta y_{t+1} - H_t(x_t,y_t)\right]\\
&= \inf_{x\in\N^p}E\sum_{t=0}^T\left[-x_t\cdot E_t[\Delta y_{t+1}] - H_t(x_t,y_t)\right]\\
&= E\sum_{t=0}^T\inf_{x_t\in\reals^{n_t}}\left[-x_t\cdot E_t[\Delta y_{t+1}] - H_t(x_t,y_t)\right]\\
&= -E\sum_{t=0}^T\sup_{x_t\in\reals^{n_t}}\sup_{v_t\in\reals^d}\left[x_t\cdot E_t[\Delta y_{t+1}] + v_t\cdot y_t-L_t(x_t,v_t)\right]\\
&= -E\sum_{t=0}^TL^*_t(E_t[\Delta y_{t+1}],y_t).
\end{align*}
The dual problem thus looks much like the primal except that the (forward) difference term enters the integral functional through the conditional expectation.
\end{example}


The above formulation of the Bolza problem was inspired by its continuous-time analogs. In the present discrete-time setting, the primal objective can be written as
\[
\minimize_{x\in\N}\quad E\sum_{t=0}^T\tilde L_t(x_t(\omega),x_{t-1}(\omega),\omega)
\]
for the normal integrands $\tilde L_t(x_t,x_{t-i},\omega):=L(x_t,x_t-x_{t-1},\omega)$. This format covers the stochastic extensions of the von Neumann-Gale model studied e.g.\ in Dempster, Evstigneev and Taksar~\cite{det6}.

\section{Examples from mathematical finance}\label{sec:mf}

Convex duality has long been an integral part of mathematical finance. The case of American options was already discussed in Remark~\ref{rem:sim} above. Perhaps the most famous instance is the ``fundamental theorem of asset pricing'' which, in perfectly liquid market models, relates the existence of an arbitrage opportunity with that of an equivalent martingale measure for the underlying price process; see Delbaen and Schachermayer~\cite{ds6} for a comprehensive treatment of the perfectly liquid case and Kabanov and Safarian~\cite{ks9} for extensions to markets with proportional transaction costs. Other instances of convex duality  can be found in problems of portfolio optimization or optimal consumption; see e.g.\ Cvitanik and Karatzas~\cite{ck92}, Kramkov and Schachermayer~\cite{ks99} or Karatzas and \u{Z}itkovi\'c~\cite{kz3}. Biagini~\cite{bia9} reviews utility maximization in perfectly liquid market models. Klein and Rogers~\cite{kr7} propose an abstract duality framework that unifies several earlier ones on optimal investment and consumption under market frictions. Several instances of convex duality in the financial context can be found in F\"ollmer and Schied~\cite{fs4} who give a comprehensive treatment of the classical perfectly liquid market model in finite discrete time; see Example~\ref{ex:cd} below.

We will show that, in finite discrete time, many duality frameworks in mathematical finance are instances of the abstract duality framework of Section~\ref{sec:cd}. Moreover, our framework  allows for various generalizations of existing financial models. We will study financial problems by following Kabanov~\cite{kab99} in that none of the assets is given the special role of a numeraire. Instead, all traded securities are treated symmetrically and contingent claims, consumption etc.\ take their values in the space of portfolios. This setting covers more traditional models where trading costs and claims are measured in cash; see Example~\ref{ex:cd} below.

Consider a market where $d$ securities are traded over finite discrete time $t=0,\ldots,T$. At each time $t$ and state $\omega\in\Omega$, the market is described by two closed convex sets, $C_t(\omega)\subset\reals^J$ and $D_t(\omega)\subset\reals^J$ both of which contain the origin. The set $C_t(\omega)$ consists of the portfolios that are freely available in the market at time $t$ and $D_t(\omega)$ consists of the portfolios that the investor is allowed to hold over the period $[t,t+1)$. For each $t$, the sets $C_t$ and $D_t$ are assumed to be $\F_t$-measurable. If $C_t(\omega)$ are polyhedral cones and $D_t(\omega)\equiv\reals^d$ (no portfolio constraints), we recover the model of \cite{kab99}.

A {\em contingent claim process (with physical delivery)} is a financial contract specified by an adapted $\reals^J$-valued process $u=(u_t)_{t=0}^T$. At each $t=0,\ldots,T$, the seller of the claim delivers a (possibly state dependent) portfolio $u_t$ to the buyer. Traditionally, financial mathematics has studied contingent claims that have only one payout date. This corresponds to $u_t=0$ for $t<T$. In real markets with portfolio constraints, it is important to distinguish between payments that occur at different points in time. We refer the reader to \cite{pen10,pen10b} for further discussion of the topic in the case of claims with cash-delivery. 

A trading strategy $x\in\N$ {\em superhedges} a claim process $u\in\N$ if 
\begin{equation}\label{sh}
\Delta x_t + u_t\in C_t,\ x_t\in D_t,\ t=0,\ldots,T,\ x_T=0
\end{equation}
almost surely. Here and in what follows, we always set $x_{-1}=0$. Superhedging is the basis of many results in financial mathematics. Even if superhedging is not quite feasible in many practical situations, it turns out to be a useful notion in studying more realistic approaches based on risk preferences.

\begin{example}[Cash delivery]\label{ex:cd}
Most contingent claims in practice give payments in cash. If cash is represented by the asset indexed $0$, a claim with cash delivery has $u_t=(u^0_t,0,\ldots,0)$ where $u^0$ is a scalar process. In this case, it is convenient to specify the market model by
\begin{align*}
C_t(\omega) &= \{(z^0,z)\in\reals^d\,|\, z^0+S_t(z,\omega)\le 0\},
\end{align*}
where the function $S_t(z,\omega)$ represents the cost in cash of buying a portfolio $z\in\reals^{d-1}$ at time $t$ in state $\omega$. The set $C_t$ is $\F_t$-measurable as soon as $S_t$ is an $\F_t$-measurable normal integrand on $\reals^{d-1}\times\Omega$. Such a model with portfolio constraints has been studied in \cite{pen10,pen10b}. The budget constraint $\Delta x_t + u_t\in C_t$ can now be written as 
\[
\Delta z^0_t + S_t(\Delta z_t) + u^0_t\le 0. 
\]

If there are no constraints on $z^0$ (the position on the cash account), we can substitute out the variables $z^0_t$ for $t=1,\ldots,T$ to write the superhedging condition as
\begin{equation}\label{budget-cd}
\sum_{t=0}^Tu^0_t + \sum_{t=0}^TS_t(\Delta z_t)\le z^0_0;
\end{equation}
see \cite[Example~3.1]{pen10}. In this setting, there is no need to discriminate between payments at different points in time. If, moreover, the functions $S_t(\cdot,\omega)$ are linear so that $S_t(z,\omega)=s_t(\omega)\cdot z$ for an adapted price process $s=(s_t)$, we can rearrange terms in \eqref{budget-cd} to write it in terms of a ``stochastic integral'' as
\[
\sum_{t=0}^Tu^0_t\le z^0_0 + \sum_{t=0}^T z_{t-1}\cdot\Delta s_t.
\]
This is the traditional formulation of the superhedging problem. It is based on the assumptions that, the contingent claim gives payments in terms of a perfectly liquid asset and that assets can be traded without a cost. In practice, however, neither of the assumptions holds.
\end{example}

The following example gives a dual characterization of the set of claims that can be superhedged at zero cost. Such characterizations are used in various results in mathematical finance; see Examples~\ref{ex:psh} and \ref{ex:arb} below.

\begin{example}[Consistent price systems]\label{ex:cps}
The superhedging condition \eqref{sh} can be studied in our general duality framework with $n_t=d$, $m=(T+1)d$ and 
\[
f(x,u,\omega) = 
\begin{cases}
0 & \text{if $\Delta x_t + u_t\in C_t(\omega)$, $x_t\in D_t(\omega)$, $x_T=0$}\\
+\infty & \text{otherwise}.
\end{cases}
\]
Indeed, we then get $\varphi=\delta_\C$, where
\[
\C = \{u\in L^p\,|\,\exists x\in\N:\ \Delta x_t + u_t\in C_t,\ x_t\in D_t,\ x_T=0\}
\]
is the set of (not necessarily adapted) claim processes that can be superhedged at zero cost. This fits the framework of Example~\ref{ex:bolza} with 
\[
L_t(x,u,\omega)=
\begin{cases}
0 & \text{if $x\in D_t(\omega)$ and $u\in C_t(\omega)$},\\
+\infty & \text{otherwise},
\end{cases}
\]
where  $D_T(\omega):=\{0\}$. Since 
\[
L^*(v,y,\omega) =  \sigma_{D_t(\omega)}(v) + \sigma_{C_t(\omega)}(y),
\]
we get for every $y\in\N^q$
\[
g(y) = -E\sum_{t=0}^T[\sigma_{D_t}(E_t\Delta y_{t+1}) + \sigma_{C_t}(y_t)],
\]
where $\sigma_{D_T}= 0$. In the unconstrained case where $D_t=\reals^d$ for $t=0,\ldots,T-1$, we have $\sigma_{D_t}=\delta_{\{0\}}$ so that $g(y)=-\infty$ unless $y$ is a martingale and thus, we recover \cite[Lemma~4.3]{pp10}. 
When $C$ and $D$ are conical, we have $g=-\delta_\D$, where
\[
\D=\{y\in\N^q\,|\, E_t\Delta y_{t+1}\in D^*_t,\ y_t\in C^*_t\}
\]
and $C^*_t(\omega)$ and $D^*_t(\omega)$ are the polar cones of $C_t(\omega)$ and $D_t(\omega)$ respectively. The elements of $\D$ are called {\em consistent price systems} for the market model $(C,D)$. The notion of a consistent price system was introduced in Kabanov~\cite{kab99} for the case of a polyhedral conical $C$ and $D\equiv\reals^d$.
\end{example}

Much of trading in financial markets consists of exchanging sequences of cash-flows. In a typical situation, one exchanges a claim process $u\in\N^p$ for a multiple of another claim process $p\in\N^p$ -- the {\em premium process}. Traditionally, financial mathematics has been mainly concerned with the special case where $p_t=0$ for $t>0$ and $u_t=0$ for $t<T$. The best known application of this special setting is the pricing of European options. Due to portfolio constraints, however, premiums as well as claims are often paid over multiple points in time. Examples include swap contracts as well as various insurance contracts where premium payments are made throughout the life of the contract.

The {\em superhedging cost} of a claim process $u\in\N^p$ in terms of a premium process $p\in\N^p$ is defined as 
\[
\varphi(u)=\inf\{\alpha\,|\,u-\alpha p\in\C\},
\]
where $\C$ is the set of claim processes that can be superhedged with zero cost; see Example~\ref{ex:cps}. The special case where claims and premiums are paid in cash has been studied in \cite{pen10b}. The following addresses the general case of ``physical delivery''.

\begin{example}[Pricing by superhedging]\label{ex:psh}
The superhedging cost is the value function in our general framework with
\[
f(x,u,\omega) = 
\begin{cases}
\alpha & \text{if $\Delta z_t + u_t-\alpha p_t\in C_t(\omega)$, $z_t\in D_t(\omega)$, $z_T=0$}\\
+\infty & \text{otherwise},
\end{cases}
\]
where $x_0 = (z_0,\alpha)$ and $x_t = z_t$ for $t=1,\ldots,T$. We have assumed for simplicity that $\F_0=\{\emptyset,\Omega\}$ so that $z_0$ is deterministic. Alternatively, we could introduce a new decision stage at time $t=-1$ with $x_{-1}=\alpha$ and $\F_{-1}=\{\emptyset,\Omega\}$. We get
\begin{align*}
l(x,y,\omega) &=\inf_{u\in\reals^m}\{f(x,u,\omega)-u\cdot y\}\\
&= \inf_{u\in\reals^m}\left\{\left.\alpha - u\cdot y\,\right|\,\Delta z_t+u_t-\alpha p_t\in C_t(\omega),\ z_t\in D_t(\omega),\ z_T=0\right\}\\
&=\inf_{z\in\reals^m}\left\{\left.\alpha+\sum_{t=0}^T(\Delta z_t-w_t-\alpha p_t)\cdot y_t\,\right|\,w_t\in C_t(\omega),\ z_t\in D_t(\omega),\ z_T=0\right\}\\
&=
\begin{cases}
\alpha + \sum_{t=0}^T(\Delta z_t-\alpha p_t)\cdot y_t - \sum_{t=0}^T\sigma_{C_t(\omega)}(y_t) & \text{if $z_t\in D_t(\omega)$ and $z_T=0$},\\
+\infty & \text{otherwise}
\end{cases}\\
&=
\begin{cases}
\alpha(1-\sum_{t=0}^Tp_t\cdot y_t)-\sum_{t=0}^{T-1}z_t\cdot\Delta y_{t+1} - \sum_{t=0}^T\sigma_{C_t(\omega)}(y_t) & \text{if $z_t\in D_t(\omega)$},\\
+\infty & \text{otherwise}.
\end{cases}
\end{align*}
This satisfies the assumptions of Theorem~\ref{thm} with $r=\infty$ so 
\begin{align*}
g(y) &= \inf_{x\in\N^\infty}\left\{\left.E\left[\alpha(1-\sum_{t=0}^Tp_t\cdot y_t)-\sum_{t=0}^{T-1}z_t\cdot\Delta y_{t+1} - \sum_{t=0}^T\sigma_{C_t(\omega)}(y_t)\right]\,\right|\,z_t\in D_t\right\}\\
&= 
\begin{cases}
\inf_{z\in\N^\infty}\left\{-E\left[\left.\sum_{t=0}^{T-1}z_t\cdot E_t\Delta y_{t+1} + \sum_{t=0}^T\sigma_{C_t}(y_t)\right]\right|\,z_t\in D_t\right\} & \text{if $E\sum_{t=0}^Tp_ty_t=1$}\\
-\infty & \text{otherwise}
\end{cases}\\
&=
\begin{cases}
-E\left[\sum_{t=0}^{T-1}\sigma_{D_t(\omega)}(E_t\Delta y_{t+1}) + \sum_{t=0}^T\sigma_{C_t(\omega)}(y_t)\right] & \text{if $E\sum_{t=0}^Tp_t\cdot y_t = 1$},\\
-\infty & \text{otherwise}.
\end{cases}
\end{align*}
where the last equality comes from the interchange rule in Theorem~\ref{int}. This expression corresponds to \cite[Lemma~7.1]{pen10} which addressed contingent claim processes with cash delivery. By Jensen's inequality, 
\[
E_t\sigma_{C_t(\omega)}(y_t)\ge \sigma_{C_t(\omega)}(E_ty_t),
\]
so that $g(y)\le g(\pi y)$, where $\pi$ denotes the ``projection'' $(y_t)_{t=0}^T\mapsto(E_ty_t)_{t=0}^T$. This implies that, for adapted claims $u\in \N^p$,
\begin{align*}
(\cl\varphi)(u) &= \sup_{y\in L^q}\{\langle u,y\rangle +g(y)\}\\
&= \sup_{y\in \N^q}\{\langle u,y\rangle +g(y)\},
\end{align*}
where $\N^q$ denotes the set of {\em adapted} $p$-integrable processes $y=(y_t)_{t=0}^T$. This corresponds to \cite[Theorem~10]{pen10b} on claims with cash-delivery. Closedness conditions will be given in  Corollary~\ref{cor:shcl} below. If $C$ and $D$ are conical, the above formula can be written as
\[
(\cl\varphi)(u) = \sup_{y\in\D}\left\{\left.E\sum_{t=0}^Tu_t\cdot y_t\,\right|\,E\sum_{t=0}^Tp_t\cdot y_t = 1\right\}.
\]
where $\D$ is the set of consistent price systems defined in Example~\ref{ex:cps}.
\end{example}

In the case of classical perfectly liquid markets, we can write the above results in a more familiar form.

\begin{example}[Martingale measures]\label{ex:mm}
Consider Example~\ref{ex:cd} in the classical perfectly liquid case, where $D\equiv\reals^d$ and
\[
C_t(\omega) = \{(x^0,x^1)\,|\, x^0+s_t(\omega)\cdot x^1\le 0\}.
\]
We get $C_t(\omega)^*=\{(y^0,y^1)\,|\,y^0\ge 0,\ y^1 = s_t(\omega)y^0\}$, so the set of consistent price systems becomes
\[
\D=\{y\in\N^q\,|\,E_t\Delta y_{t+1}=0,\ y^0\ge 0,\ y^1_t = s_ty^0_t\}.
\]
Consider now Example~\ref{ex:psh} in the case where $p_t=0$ for $t=1,\ldots,T$ and $p_0=(1,0,\ldots,0)$. This corresponds to the classical pricing problem where the claim $u$ is exchanged for a cash payment at time $t=0$. The condition $E\sum_{t=0}^Tp_t\cdot y_t = 1$ now means that $y^0_0=1$. If $u$ is a claim process with cash delivery, i.e.\ $u_t=(u^0_t,0)$, we get
\begin{align*}
(\cl\varphi)(u) &= \sup_{y\in\N^q}\left\{\left.E\sum_{t=0}^Tu^0_ty^0_t\,\right|\, E_t\Delta y_{t+1}=0,\ y^0\ge 0,\ y^1_t = s_ty^0_t,\ y^0_0=1\right\}\\
&= \sup_{Q\in\M^q(s)}E^Q\sum_{t=0}^Tu^0_t,
\end{align*}
where $\M^q(s)$ denotes the set of probability measures $Q$ under which the price process $s$ is a martingale and whose density $dQ/dP$ is $q$-integrable. Indeed, it follows from the law of iterated expectations that the densities of such measures correspond to the random variables $y^0_T$ above. For claims with $u^0_t=0$ for $t<T$ we obtain the classical dual representation of the superhedging cost. See \cite{pen10b} for further discussion and references.
\end{example}

We end this section with a model of optimal consumption problems in the general illiquid market model.

\begin{example}[Optimal consumption]\label{ex:oc}
Consider the problem
\begin{alignat*}{2}
&\maximize_{x,c\in\N}\quad& E\sum_{t=0}^T&U_t(c_t)\\
&\st \quad & \Delta x_t + c_t & \in C_t,\quad x_t\in D_t\quad t=0,\ldots,T,
\end{alignat*}
where $D_T:=\{0\}$ and $U_t$ is an $\F_t$-measurable concave normal integrand on $\reals^d\times\Omega$. This represents a problem of optimal consumption where possibly all traded assets can be directly consumed. To model situations where some of the assets cannot be consumed, one can set $U_t(c,\omega)=-\infty$ for $c$ outside of the feasible consumption set. Defining $\C$ as in Example~\ref{ex:cps}, we can write the problem concisely as 
\[
\maximize\quad E\sum_{t=0}^TU_t(c_t)\quad\text{{\rm over}}\ c\in\C.
\]
This is the primal problem of Example~\ref{ex:bolza} in the case
\[
L_t(x,v,\omega) =
\begin{cases}
\inf_{c_t\in\reals^d}\{-U_t(c_t,\omega) \,|\, v_t + c_t\in C_t(\omega)\} & \text{if $x_t\in D_t(\omega)$},\\
+\infty & \text{otherwise}.
\end{cases}
\]
By \cite[Proposition~14.47]{rw98}, $L_t$ is an $\F_t$-measurable normal integrand as soon as it is lower semicontinuous in $(x,v)$. Conditions for lower semiconinuity, in turn, can be obtained by pointwise application of \cite[Theorem~9.2]{roc70a}. It is easily checked that
\[
L_t^*(v,y,\omega) = \sigma_{D_t(\omega)}(v) + \sigma_{C_t(\omega)}(y) - U_t^*(y,\omega),
\]
where
\[
U_t^*(y,\omega) = \inf_{c\in\reals^d}\{c\cdot y-U_t(c,\omega)\}
\]
is the conjugate of $U_t$ in the concave sense. If $C$ and $D$ are conical, we get
\[
g(y) = \begin{cases}
E\sum_{t=0}^TU^*_t(y_t) & \text{if $y\in\D$},\\
-\infty & \text{otherwise},
\end{cases}
\]
where $\D$ is the set of consistent price systems defined in Example~\ref{ex:cps}. The dual problem can then be written in the symmetric form
\[
\maximize\quad E\sum_{t=0}^TU^*_t(y_t)\quad\text{{\rm over}}\ y\in\D.
\]
The dual pair of optimization problems above can be seen as a generalization (in discrete time) of the optimal consumption duality framework of Karatzas and \u{Z}itkovi\'c~\cite{kz3} where the numeraire asset was consumed in a perfectly liquid market model in continuous time.
\end{example}

\section{Some closedness criteria}\label{sec:cl}





Much of duality theory in convex analysis has been concerned with optimality conditions and the attainment of dual optimum. Dual attainment is equivalent to the subdifferentiability of the value function $\varphi$ at the origin, which in turn is implied by continuity; see \cite[Section~7]{roc74}. In operations research, several ``constraint qualifications'' have bee proposed to guarantee the continuity of $\varphi$ at the origin. Unfortunately, such conditions fail in many infinite dimensional applications. In order to get the mere absence of a duality gap, it is sufficient (as well as necessary) that $\varphi$ be proper and lower semicontinuous at the origin. In this section, we outline techniques for establishing the lower semicontinuity of $\varphi$ and the attainment of the primal optimum.

The traditional technique for achieving lower semicontinuity of $\varphi$ would be to introduce a topology on (an appropriate subspace of) $\N$, to show that $I_f$ is lower semicontinuous and to impose inf-compactness conditions on $I_f$ with respect to $x$; see e.g.\ \cite[Theorem~1.17]{rw98}. This is essentially the  ``direct method'' in calculus of variations for verifying the existence of a solution to a minimization problem; see e.g.\ \cite{abm5}. As long as the topology is strong enough to imply almost sure convergence of converging sequences, the lower semicontinuity of $I_f$ often follows from Fatou's lemma and pointwise lower semicontinuity of normal integrands. The inf-compactness property, on the other hand, is often obtained with Alaoglu-type arguments provided the topology is weak enough. In particular, $\varphi$ is $\sigma(L^\infty,L^1)$-closed when the feasible set is $L^\infty$-bounded locally uniformly in $u$. This result applies already to many problems arising in practice and, in particular, to the optimal stopping problem in Example~\ref{ex:os}.

In some applications, the compactness condition does not hold. In the convex setting, the following version of a theorem of Koml\'os'~\cite{kom67} can often be used as a substitute.

\begin{lemma}[Koml\'os' theorem]\label{lem:kom}
Let $(x^\nu)_{\nu=1}^\infty$ be a sequence in $L^0(\Omega,\F,P;\reals^n)$ which is almost surely bounded in the sense that 
\[
\sup_\nu|x^\nu(\omega)|<\infty\quad P\text{-a.s.}
\]
Then there is a sequence of convex combinations $\bar x^\nu\in\co\{x^\mu\,|\,\mu\ge\nu\}$ that converges almost surely to an $\reals^n$-valued function.
\end{lemma}

\begin{proof}
See Delbaen and Schachermayer~\cite{ds6} or Kabanov and Safarian~\cite{ks9}.
\end{proof}

Different versions of the Koml\'os' theorem have long been used in calculus of variations; see e.g.\ Balder~\cite{bal89} or Schachermayer~\cite{sch92} for an application to the classical perfectly liquid market model in Example~\ref{ex:mm}.

The almost sure boundedness in Lemma~\ref{lem:kom} can sometimes be obtained by pointwise application of classical finite-dimensional boundedness conditions on directions of recession. Given a convex set $C$, we will denote its {\em recession cone} by
\[
C^\infty = \{z\,|\,x+\alpha z\in C,\ \forall x\in C,\ \alpha>0\}.
\]
By \cite[Theorem~8.4]{roc70a}, a closed convex set $C$ in a finite-dimensional space is bounded if and only if $C^\infty=\{0\}$. The proof of this result is based on the classical Bolzano-Weierstrass theorem on converging subsequences in finite-dimensional spaces. The following simple modification of \cite[Lemma~2]{ks1b} generalizes the finite-dimensional Bolzano-Weierstrass theorem to the present stochastic setting.

\begin{lemma}\label{rss}
For an almost surely bounded sequence $(x^\nu)_{\nu=1}^\infty$ in $\N$ there exists a strictly increasing sequence of $\F_T$-measurable integer-valued functions $(\tau^\nu)$ and an $x\in\N$ such that
\[
x^{\tau^\nu}\to x
\]
almost surely.
\end{lemma}

\begin{proof}
Applying \cite[Lemma~2]{ks1b} to $(x_0^\nu)_{\nu=1}^\infty$ we get an $\F_0$-measurable random subsequence $\tau_0^\nu$ such that $x_0^{\tau_0^\nu}\to x_0$ for an $x_0\in L^0(\Omega,\F_0,P;\reals^{n_0})$. Applying \cite[Lemma~2]{ks1b} next to $(x_1^{\tau_0^\nu})_{\nu=1}^\infty$ we get an $\F_1$-measurable subsequence $\tau_1^\nu$ of $\tau_0^\nu$ such that $x_1^{\tau_1^\nu}\to x_1$ for an $x_1\in L^0(\Omega,\F_1,P;\reals^{n_1})$. Since $x_0^{\tau_0^\nu}\to x_0$ we also have $x_0^{\tau_1^\nu}\to x_0$. Extracting further subsequences similarly for $t=2,\ldots,T$ we arrive at the conclusion.
\end{proof}


Sequences of the form $(x^{\tau^\nu})_{\nu=1}^\infty$ in the above lemma are called {\em random subsequences} of the original sequence $(x^\nu)_{\nu=1}^\infty$. 

If $C:\Omega\tos\reals^n$ is a closed convex-valued $\F$-measurable mapping, then $C^\infty(\omega):=C(\omega)^\infty$ defines an $\F$-measurable mapping whose values are closed convex cones; see \cite[Exercise~14.21]{rw98}. The following result generalizes \cite[Theorem~8.4]{roc70a} to stochastic models in finite discrete time. The proof follows the inductive argument in the proof of \cite[Theorem~3.3]{pp10} with some simplifications. Theorem~3.3 of \cite{pp10} deals with Example~\ref{sh} in the case $D_t\equiv\reals^d$ and its proof builds on earlier techniques developed for conical models of financial markets; see e.g.\ \cite{sch4} or \cite{ks9}.

\begin{theorem}\label{thm:bdd}
Let $C:\Omega\tos\reals^n$ be closed convex-valued and $\F$-measurable. Every sequence in the set $\C=\{x\in\N\,|\,x\in C\ a.s.\}$ is almost surely bounded if and only if $\{x\in\N\,|\,x\in C^\infty\ a.s.\}=\{0\}$.
\end{theorem}

\begin{proof}
If the recession condition fails, then $\C$ contains a half-line so it cannot be a.s.\ bounded. To prove the converse, we may assume that $0\in C$ almost surely. Indeed, if $\C$ is empty, there is nothing to prove. Otherwise, we take any $x\in\C$, set $C(\omega):=C(\omega)-x(\omega)$ and note that the translation does not affect the recession cone of $C$ or the almost sure boundedness of $\C$.

We use induction on $T$. Let $T>0$ and assume first that the claim holds for every $(T-1)$-period model. Let $(x^\nu)_{\nu=1}^\infty\subset\C$ and consider the following two complementary cases.

Case 1: $\rho(\omega):=\sup|x_0^\nu(\omega)|<\infty$ almost surely. Let 
\begin{align*}
\N_1 &:= \{(x_1)_{t=1}^T\,|\,x_t\in L^0(\Omega,\F_t,P;\reals^{n_t})\},\\
C_1(\omega) &:= \{(x_1,\ldots,x_T)\,|\,\exists x_0\in\rho(\omega)\uball:\ (x_0,\ldots,x_T)\in C(\omega)\}.
\end{align*}
By Proposition~14.11(a) and Proposition~14.13(a) of \cite{rw98}, $C_1$ is $\F$-measurable and, by \cite[Theorem~9.1]{roc70a}, it is closed and convex-valued with
\[
C_1^\infty(\omega) = \{(x_1,\ldots,x_T)\,|\, (0,x_1,\ldots,x_T)\in C^\infty(\omega)\}.
\]
Our assumption thus implies that $\{x\in\N_1\,|\, x\in C_1^\infty\ P\text{-a.s.}\} = \{0\}$ so the sequence $(x^\nu_1,\ldots,x^\nu_T)$ is almost surely bounded by the induction hypothesis.

Case 2: the set $A=\{\omega\in\Omega\,|\,\sup|x_0^\nu(\omega)|=\infty\}$ has positive probability. Let $\alpha^\nu=\chi_A/\max\{|x_0^\nu|,1\}$ and $\bar x^\nu=\alpha^\nu x^\nu$. Passing to an $\F_0$-measurable random subsequence if necessary, we may assume that $\alpha^\nu\downto 0$ almost surely. Since $\alpha^\nu$ are $\F_0$-measurable, $\bar x^\nu\in\N$. We also have that  
\[
\bar x^\nu\in\alpha^\nu C
\]
and $|\bar x_0^\nu|\le 1$ almost surely. Since $\alpha^\nu\le 1$ and $0\in C$ we get $\alpha^\nu C\subset C$, by convexity. We are thus in the same situation as in case 1, so $(\bar x^\nu)_{\nu=1}^\infty$ is almost surely bounded. By Lemma~\ref{rss}, there is an $\F_T$-measurable subsequence $\tau^\nu$ such that $(\bar x^{\tau^\nu})_{\nu=1}^\infty$ converges almost surely to an $\bar x\in\N$. By \cite[Theorem~8.2]{roc70a},
\[
\bar x\in C^\infty
\]
almost surely, so $\bar x=0$ by the assumption. This contradicts the positivity of $P(A)$ since on $A$, we have $|\bar x^\nu_0(\omega)|\upto 1$ so that  $|\bar x_0(\omega)|=1$.  

It remains to prove the claim for $T=0$. This can be done as in Case 2 above except that now we do not need to refer to Case 1 for the boundedness of $(\bar x^\nu)_{\nu=1}^\infty$ (This is essentially the finite-dimensional argument in \cite[Theorem~8.4]{roc70a}).
\end{proof}

With Theorem~\ref{thm:bdd}, we can generalize finite-dimensional closedness results to the present stochastic setting much like \cite[Theorem~8.4]{roc70a} was used in Section~9 of \cite{roc70a}. Orthogonal projections, which are central in the arguments of \cite[Section~9 ]{roc70a}, are not well-defined in the space $\N$ but the following lemma (whose origins can be traced back to Schachermayer~\cite{sch92}) can be used instead.

\begin{lemma}\label{lem:proj}
Let $L:\Omega\tos\reals^n$ be an $\F$-measurable mapping whose values are linear. For each $t=0,\ldots,T$, there is an $\F_t$-measurable linear-valued mapping $L_t:\Omega\tos\reals^{n_t}$ such that
\begin{multline*}
\{x_t\in L^0(\Omega,\F_t,P;\reals^{n_t})\,|\, \exists z\in\N:\ (0,\ldots,0,x_t,z_{t+1}\ldots,z_T)\in L\}\\
= \{x_t\in L^0(\Omega,\F_t,P;\reals^{n_t})\,|\, x_t\in L_t\ P\text{-a.s.}\}.
\end{multline*}
\end{lemma}

\begin{proof}
It suffices to prove the claim for $t=0$ since otherwise we can replace $L$ by the set-valued mapping $\bar L(\omega)=\{(x_t,\ldots,x_T)\,|\,(0,\ldots,0,x_t,\ldots,x_T)\in L(\omega)\}$ which is also closed-valued and $\F$-measurable, by Proposition~14.11(a) and Proposition~14.13(a) of \cite{rw98}. For any $\F$-measurable closed-valued mapping $S:\Omega\tos\reals^k$, there is an $\F_t$-measurable mapping $\Gamma_tS$ whose $\F_t$-measurable selectors coincide with those of $S$. Indeed, it suffices to check that the proof of \cite[Theorem~3.1]{hu77} (and the proofs of the lemmas used in it) goes through in the case $p=0$ with the norm replaced by the metric $d(x^1,x^2)=E\min\{|x^1(\omega)-x^2(\omega)|,1\}$. We define $L_t$ recursively by $L_T=\Gamma_TL$ and
\[
L_t=\Gamma_tP_tL_{t+1},
\]
where $(P_tL_{t+1})(\omega):=\{(x_0,\ldots,x_t)\,|\, \exists x_{t+1}\in\reals^{n_{t+1}}:\ (x_0,\ldots,x_{t+1})\in L_{t+1}(\omega)\}$. The mapping $P_tL_{t+1}$ is $\F_{t+1}$-measurable (see \cite[Proposition~14.13(a)]{rw98}) and linear-valued. It is clear that if $x_0$ belongs to the set on the right side then $x_0\in L_0$ almost surely. The reverse direction follows from repeated application of the theorem on measurable selections; see e.g.\ \cite[Corollary~14.5]{rw98}.
\end{proof}

The following can be seen as a generalization of \cite[Theorem~9.1]{roc70a} to our stochastic setting.

\begin{theorem}\label{thm:cl}
Let $C:\Omega\tos\reals^n\times\reals^m$ be a closed convex-valued $\F$-measurable mapping such that $\{x\in\N\,|\,(x,0)\in C^\infty\ a.s.\}$ is a linear space. Then the set 
\[
\C = \{u\in L^p\,|\,\exists x\in\N:\ (x,u)\in C\ a.s.\}
\]
is $\sigma(L^p,L^q)$-closed.
\end{theorem}

\begin{proof}
Let $L=\{x\in\reals^n\,|\, (x,0)\in C^\infty(\omega)\cap(-C^\infty(\omega))\}$ and define $L_t$ as in Lemma~\ref{lem:proj}. We have 
\[
\C = \{u\in L^p\,|\,\exists x\in\N:\ x_t\in L_t^\perp,\ (x,u)\in C\ a.s.\}.
\]
Indeed, let $x_0^0(\omega)$ be the pointwise orthogonal projection of $x_0(\omega)$ to $L_0(\omega)$. By definition of $L_t$, there is an extension $x^0\in\N$ of $x^0_0$ such that $(x^0,0)\in C^0$ almost surely. Defining $\tilde x^0=x-x^0$, we have $\tilde x^0_0\in L_0^\perp$ and $(\tilde x^0,u)\in C$ almost surely. Repeating the procedure for $t=1,\ldots,T$ we arrive at an $\tilde x^T\in\N$ with $\tilde x^T_t\in L_t^\perp$ and $(\tilde x^T,u)\in C$ almost surely.

Assume first that $p=1$. Since $\C$ is convex it suffices to prove $\tau(L^1,L^\infty)$-closedness. Since $\tau(L^1,L^\infty)$ is the norm topology, it suffices to verify sequential closedness. So assume that $(u^\nu)_{\nu=1}^\infty\subset\C$ converges to a $u$ in norm and let $x^\nu\in\N$ be such that $x^\nu_t\in L_t^\perp$ and $(x^\nu,u^\nu)\in C$. Passing to a subsequence, we may assume that $u^\nu\to u$ almost surely, so that the measurable function $\rho(\omega) := \sup_\nu|u^\nu(\omega)|$ is almost surely finite. Each $(x^\nu,u^\nu)$ thus belongs to the set
\[
\C_\rho=\{(x,u)\in\N\times L^0\,|\, (x_t,u_t)\in C_\rho\ \text{a.s.}\},
\]
where $C_\rho(\omega) = \{(x,u)\,|\, x_t\in L_t^\perp(\omega),\ u\in\rho(\omega)\uball,\ (x,u)\in C(\omega)\},$
By \cite[Corollary~8.3.3]{roc70a}, 
\[
C_\rho^\infty(\omega)=\{(x,0)\,|\, x_t\in L_t^\perp(\omega),\ (x,0)\in C^\infty(\omega)\},
\]
so, by the linearity assumption,
\begin{align*}
\{(x,u)\in\N&\times L^0\,|\, (x,u)\in C_\rho^\infty\ \text{a.s.}\}\\
&= \{(x,0)\in\N\times L^0\,|\, x_t\in L_t^\perp,\ (x,0)\in C^\infty\cap(-C^\infty)\ \text{a.s.}\}\\
&= \{x\in\N\,|\,x_t\in L_t^\perp,\ x\in L_t\ \text{a.s.}\}\times\{0\},
\end{align*}
which equals $\{0,0\}$, by the definition of $L_t$. By Theorem~\ref{thm:bdd}, the sequence $(x^\nu,u^\nu)_{\nu=1}^\infty$ is then almost surely bounded. By Lemma~\ref{lem:kom}, there is a sequence of convex combinations $(\bar x^\nu,\bar u^\nu)_{\nu=1}^\infty$ that converges almost surely to a point $(\bar x,\bar u)$. We have $\bar u\in\C$ since $C$ is convex and closed-valued and $\bar u=u$ since the original sequence $(u^\nu)_{\nu=1}^\infty$ was convergent to $u$. 

Now let $p\in[1,\infty]$ be arbitrary. We have $\C = \{u\in L^p\,|\,Au\in\C^1\}$, where $\C^1$ denotes the set $\C$ in the case $p=1$ considered above and $A:(L^p,\sigma(L^p,L^q))\to(L^1,\sigma(L^1,L^\infty))$ is the natural injection. Since $A$ is continuous, the $\sigma(L^p,L^q)$-closedness of $\C$ follows from the $\sigma(L^1,L^\infty)$-closedness of $\C^1$.
\end{proof}

The above result can be seen as a lower-semicontinuity result for the value function $\varphi$ is situations where it takes the form of an indicator function. Theorem~\ref{thm:bdd} and Lemma~\ref{lem:proj} allow the verification of the lower-semicontinuity of $\varphi$ in more general situations as well. This will be the subject of a separate article. We end this paper by showing how Theorem~\ref{thm:cl} yields some fundamental results in financial mathematics. An early application of recession analysis to portfolio optimization can be found in Bertsekas~\cite{ber74}.

\begin{example}[The no-arbitrage condition]\label{ex:arb}
Consider the market model studied in Section~\ref{sec:mf}. The model is said to satisfy the {\em no arbitrage} condition if
\begin{equation}\label{na}
\C\cap\N_+=\{0\},
\end{equation}
where $\C$ is the set of claim processes that can be superhedged without a cost (see Example~\ref{ex:cps}) and $\N_+$ is the set of nonnegative adapted processes. The no-arbitrage condition means that it is not possible to superhedge nontrivial nonnegative claims by costless transactions in the financial market. Condition \eqref{na} was studied in \cite{pp10}, where it was related to more traditional formulations of the no-arbitrage condition. In particular, in the classical perfectly liquid market model (see Example~\ref{ex:mm}) with strictly positive market prices, \eqref{na} is equivalent to the classical no-arbitrage condition expressed in terms of claims with cash-delivery and a single payout date; see e.g.\ \cite{ds6}. In markets with portfolio constraints, however, the above formulation in terms of claims with multiple payout dates is more meaningful; see \cite{pen10,pen10b}.

It was shown in Schachermayer~\cite[Section~2]{sch92} that, in case of the unconstrained linear market model in Example~\ref{ex:mm}, the no-arbitrage condition implies that $\C$ is closed. Theorem~\ref{thm:cl} yields a simple proof of this important result. Indeed, we now have
\[
C(\omega)=\{(x,u)\in\reals^n\times\reals^m\,|\,\Delta x_t+u_t\in C_t(\omega),\ x_T=0\},
\]
where $C_t(\omega) = \{(x^0,x^1)\,|\, x^0+s_t(\omega)\cdot x^1\le 0\}$. Since $C(\omega)$ is conical, we have $C^\infty(\omega)=C(\omega)$ and the condition in Theorem~\ref{thm:cl} can be written as
\[
\{x\in\N\,|\,\Delta x_t\in C_t,\ x_T=0\ a.s.\} = \{x\in\N\,|\,\Delta x_t\in C^0_t,\ x_T=0\ a.s.\},
\]
where $C^0_t(\omega) := C_t(\omega)\cap(-C_t(\omega))$. If this condition fails, there is an $x\in\N$ such that $\Delta x_t\in C_t$ and $x_T=0$ almost surely but for some $t$ and a set $A\in\F_t$ of positive probability, $\Delta x_t\notin C^0_t$. Since $C_t(\omega)$ is a half-space, we have $C_t(\omega)\setminus C_t^0(\omega)=\inte C_t(\omega)$. Given a nonzero vector $e\in\reals^d_+$, the $\F_t$-measurable nonnegative variable
\[
\varepsilon(\omega):=\max\{\alpha\,|\,\Delta x_t(\omega)+\alpha e\in C_t(\omega)\}
\]
is thus strictly positive on $A$. We then have that $x$ superhedges the nontrivial claim process defined by $u_t(\omega)=\varepsilon(\omega)e$ and $u_s=0$ for $s\ne t$. This violates the no-arbitrage condition \eqref{na}, so the closedness condition of Theorem~\ref{thm:cl} must hold under \eqref{na}.

The above argument extends directly to market models with transaction costs (without portfolio constraints) provided one slightly strengthens the no-arbitrage condition; see~\cite{sch4,ks9} for conical polyhedral market models or \cite{pp10} for a more general convex model.
\end{example}

The closedness of $\C$ combined with the Kreps-Yan theorem (see e.g.\ \cite{jns5}), yields the following famous result of Dalang, Morton and Willinger~\cite{dmw90}. The proof of the Kreps-Yan theorem is based on separation and exhaustion arguments.

\begin{corollary}[Fundamental theorem of asset pricing]
The perfectly liquid market model of Example~\ref{ex:mm} with a strictly positive market price process $s$ satisfies the no arbitrage condition if and only if there is a probability measure equivalent to $P$ under which $s$ is a martingale.
\end{corollary}

\begin{proof}
  Assume that the no-arbitrage condition holds. By Example~\ref{ex:arb}, $\C$ is closed in $\N^p$. The Kreps-Yan theorem then gives the existence of a strictly positive $y\in\N^q$ such that $y\in\C^*$, where $\C^*$ is the polar cone of $\C$. It was shown in Example~\ref{ex:cps} that in the case of conical market models, $\C^*$ equals the set $\D$ of consistent price systems, which in the case of perfectly liquid market models may be identified with martingale measures for $s$; see Example~\ref{ex:mm}. Similarly, strictly positive price systems $y\in\D$ correspond to martingale measures which are equivalent to~$P$.

On the other hand, the existence of a strictly positive martingale measure means that there is a strictly positive $y\in\C^*$. Any nonzero $u\in\C\cap\N_+$ would then satisfy both $E(u\cdot y)\le 0$ and $E(u\cdot y)>0$, which is clearly impossible. 
\end{proof}

The following gives closedness conditions for Example~\ref{ex:psh}. 

\begin{corollary}[Superhedging cost]\label{cor:shcl}
Consider Example~\ref{ex:psh} and assume that
\[
\{x\in\N\,|\,\Delta x\in C_t^\infty,\ x_t\in D_t^\infty,\ x_T=0\ a.s.\}
\]
is a linear space and that the premium process $p\in\N^p$ is such that $\varphi(0)>-\infty$. Then the superhedging cost
\[
\varphi(u) = \inf\{\alpha\,|\,u-\alpha p\in\C\}
\]
is closed.
\end{corollary}

\begin{proof}
The set $\C$ of claim processes that can be superhedged at zero cost corresponds to Theorem~\ref{thm:cl} with
\[
C(\omega) = \{(x,u)\in\reals^n\times\reals^m\,|\, \Delta x_t+u_t\in C_t(\omega),\ x_t\in D_t(\omega),\ x_T=0\}.
\]
By \cite[Corollary~8.3.3]{roc70a}, 
\[
C^\infty(\omega) = \{(x,u)\in\reals^n\times\reals^m\,|\, \Delta x_t+u_t\in C_t^\infty(\omega),\ x_t\in D_t^\infty(\omega),\ x_T=0\},
\]
so our assumption means that the closedness condition in Theorem~\ref{thm:cl} holds. The set $\C$ is thus closed and, in particular, algebraically (or radially) closed. The closedness of $\varphi$ then follows exactly like in the proof of \cite[Theorem~10]{pen10b}.
\end{proof}

As shown in Example~\ref{ex:arb}, the first condition coincides with the classical no-arbitrage condition in the case of the perfectly liquid market model of Example~\ref{ex:mm}. The condition generalizes the closedness conditions given in \cite[Theorem~18]{pen10b} and in Theorem~13 of Kreher~\cite{dor9}. Relations of the first condition to certain generalized no-arbitrage conditions have been studied in \cite[Section~7]{dor9}. The condition on the premium process $p$ is mild. It means that the premium is a contingent claim that is not freely available in the market at unlimited amounts; see \cite{pen10b} for further discussion. 

More results on the lower semicontinuity of the value function as well as extensions of the presented duality framework to a continuous-time setting will be presented in separate articles.

\bibliographystyle{plain}
\bibliography{sp}

\begin{thebibliography}{10}

\bibitem{abm5}
H.~Attouch, G.~Buttazzo, and G.~Michaille.
\newblock {\em Variational Analysis in Sobolev and BV Spaces: Applications to
  PDEs and Optimization}, volume~6 of {\em MPS/SIAM Series on Optimization}.
\newblock Society for Industrial and Applied Mathematics (SIAM), Philadelphia,
  PA, 2005.

\bibitem{bp87}
K.~Back and S.~R. Pliska.
\newblock The shadow price of information in continuous time decision problems.
\newblock {\em Stochastics}, 22(2):151--186, 1987.

\bibitem{bal89}
E.~J. Balder.
\newblock Infinite-dimensional extension of a theorem of {K}oml\'os.
\newblock {\em Probab. Theory Related Fields}, 81(2):185--188, 1989.

\bibitem{bea55}
E.~M.~L. Beale.
\newblock On minimizing a convex function subject to linear inequalities.
\newblock {\em J. Roy. Statist. Soc. Ser. B.}, 17:173--184; discussion,
  194--203, 1955.
\newblock (Symposium on linear programming.).

\bibitem{ber74}
D.~P. Bertsekas.
\newblock Necessary and sufficient conditions for existence of an optimal
  portfolio.
\newblock {\em Journal of Economic Theory}, 8(2):235--247, 1974.

\bibitem{bia9}
S.~Biagini.
\newblock Expected utility maximization: the dual approach.
\newblock In R.~Cont, editor, {\em Encyclopedia of Quantitative Finance}.
  Wiley, to appear.

\bibitem{bf8}
S.~Biagini and M.~Frittelli.
\newblock A unified framework for utility maximization problems: An orlicz
  spaces approach.
\newblock {\em The Annals of Applied Probability}, 18(3):929--966, 2008.

\bibitem{ck92}
J.~Cvitani{\'c} and I.~Karatzas.
\newblock Convex duality in constrained portfolio optimization.
\newblock {\em Ann. Appl. Probab.}, 2(4):767--818, 1992.

\bibitem{dmw90}
R.~C. Dalang, A.~Morton, and W.~Willinger.
\newblock Equivalent martingale measures and no-arbitrage in stochastic
  securities market models.
\newblock {\em Stochastics Stochastics Rep.}, 29(2):185--201, 1990.

\bibitem{dan55}
G.~B. Dantzig.
\newblock Linear programming under uncertainty.
\newblock {\em Management Sci.}, 1:197--206, 1955.

\bibitem{dav92}
M.~H.~A. Davis.
\newblock Dynamic optimization: a grand unification.
\newblock In {\em Proceedings of the 31st IEEE Conference on Decision and
  Control}, volume~2, pages 2035 -- 2036, 1992.

\bibitem{db92}
M.~H.~A. Davis and G.~Burstein.
\newblock A deterministic approach to stochastic optimal control with
  application to anticipative control.
\newblock {\em Stochastics and Stochastics Reports}, 40(3\&4):203--256, 1992.

\bibitem{dk94}
M.~H.~A. Davis and I.~Karatzas.
\newblock A deterministic approach to optimal stopping.
\newblock In {\em Probability, statistics and optimisation}, Wiley Ser. Probab.
  Math. Statist. Probab. Math. Statist., pages 455--466. Wiley, Chichester,
  1994.

\bibitem{ds6}
F.~Delbaen and W.~Schachermayer.
\newblock {\em The Mathematics of Arbitrage}.
\newblock Springer Finance. Springer-Verlag, Berlin Heidelberg, 2006.

\bibitem{det6}
M.~A.~H. Dempster, I.~V. Evstigneev, and M.I. Taksar.
\newblock Asset pricing and hedging in financial markets with transaction
  costs: An approach based on the von neumann–gale model.
\newblock {\em Annals of Finance}, 2(4):327--355, 2006.

\bibitem{et76}
I.~Ekeland and R.~Temam.
\newblock {\em Convex analysis and variational problems}.
\newblock North-Holland Publishing Co., Amsterdam, 1976.
\newblock Translated from the French, Studies in Mathematics and its
  Applications, Vol. 1.

\bibitem{fs4}
H.~F{\"o}llmer and A.~Schied.
\newblock {\em Stochastic finance}, volume~27 of {\em de Gruyter Studies in
  Mathematics}.
\newblock Walter de Gruyter \& Co., Berlin, extended edition, 2004.
\newblock An introduction in discrete time.

\bibitem{gro73}
A.~Grothendieck.
\newblock {\em Topological vector spaces}.
\newblock Gordon and Breach Science Publishers, New York, 1973.
\newblock Translated from the French by Orlando Chaljub, Notes on Mathematics
  and its Applications.

\bibitem{hk4}
M.~B. Haugh and L.~Kogan.
\newblock Pricing american options: a duality approach.
\newblock {\em Oper. Res.}, 52(2):258--270, 2004.

\bibitem{hu77}
F.~Hiai and H.~Umegaki.
\newblock Integrals, conditional expectations, and martingales of multivalued
  functions.
\newblock {\em J. Multivariate Anal.}, 7(1):149--182, 1977.

\bibitem{jns5}
E.~Jouini, C.~Napp, and W.~Schachermayer.
\newblock Arbitrage and state price deflators in a general intertemporal
  framework.
\newblock {\em J. Math. Econom.}, 41(6):722--734, 2005.

\bibitem{ks9}
Y.~Kabanov and M.~Safarian.
\newblock {\em Markets with transaction costs}.
\newblock Springer Finance. Springer-Verlag, Berlin, 2009.
\newblock Mathematical theory.

\bibitem{kab99}
Yu.~M. Kabanov.
\newblock Hedging and liquidation under transaction costs in currency markets.
\newblock {\em Finance and Stochastics}, 3(2):237--248, 1999.

\bibitem{ks1b}
Yu.~M. Kabanov and Ch. Stricker.
\newblock A teachers' note on no-arbitrage criteria.
\newblock In {\em S\'eminaire de Probabilit\'es, XXXV}, volume 1755 of {\em
  Lecture Notes in Math.}, pages 149--152. Springer, Berlin, 2001.

\bibitem{kz3}
I.~Karatzas and {\u{Z}itkovi\'c}~G.
\newblock Optimal consumption from investment and random endowment in
  incomplete semimartingale markets.
\newblock {\em The Annals of Probability}, 31(4):1821–1858, 2003.

\bibitem{kr7}
I.~Klein and L.~C.~G. Rogers.
\newblock Duality in optimal investment and consumption problems with market
  frictions.
\newblock {\em Math. Finance}, 17(2):225--247, 2007.

\bibitem{kom67}
J.~Koml{\'o}s.
\newblock A generalization of a problem of {S}teinhaus.
\newblock {\em Acta Math. Acad. Sci. Hungar.}, 18:217--229, 1967.

\bibitem{ks99}
D.~Kramkov and W.~Schachermayer.
\newblock The condition on the asymptotic elasticity of utility functions and
  optimal investment in incomplete markets.
\newblock {\em Annals of Applied Probability}, 9(3):904--950, 1999.

\bibitem{dor9}
D.~Kreher.
\newblock Hedging of portfolio-valued claims under convex transaction costs and
  portfolio constraints.
\newblock Master's thesis, Humboldt-Universit\"at zu Berlin, 2009.

\bibitem{pen10}
T.~Pennanen.
\newblock Arbitrage and deflators in illiquid markets.
\newblock {\em Finance and Stochastics}, to appear.

\bibitem{pen10b}
T.~Pennanen.
\newblock Superhedging in illiquid markets.
\newblock {\em Mathematical Finance}, to appear.

\bibitem{pp10}
T.~Pennanen and I.~Penner.
\newblock Hedging of claims with physical delivery under convex transaction
  costs.
\newblock {\em SIAM Journal on Financial Mathematics}, 1:158--178, 2010.

\bibitem{roc68}
R.~T. Rockafellar.
\newblock Integrals which are convex functionals.
\newblock {\em Pacific J. Math.}, 24:525--539, 1968.

\bibitem{roc70a}
R.~T. Rockafellar.
\newblock {\em Convex analysis}.
\newblock Princeton Mathematical Series, No. 28. Princeton University Press,
  Princeton, N.J., 1970.

\bibitem{roc71}
R.~T. Rockafellar.
\newblock Integrals which are convex functionals. {II}.
\newblock {\em Pacific J. Math.}, 39:439--469, 1971.

\bibitem{roc74}
R.~T. Rockafellar.
\newblock {\em Conjugate duality and optimization}.
\newblock Society for Industrial and Applied Mathematics, Philadelphia, Pa.,
  1974.

\bibitem{roc76}
R.~T. Rockafellar.
\newblock Integral functionals, normal integrands and measurable selections.
\newblock In {\em Nonlinear operators and the calculus of variations (Summer
  School, Univ. Libre Bruxelles, Brussels, 1975)}, pages 157--207. Lecture
  Notes in Math., Vol. 543. Springer, Berlin, 1976.

\bibitem{rw76}
R.~T. Rockafellar and R.~J.-B. Wets.
\newblock Nonanticipativity and {$L^1$}-martingales in stochastic optimization
  problems.
\newblock {\em Math. Programming Stud.}, (6):170--187, 1976.
\newblock Stochastic systems: modeling, identification and optimization, II
  (Proc. Sympos., Univ Kentucky, Lexington, Ky., 1975).

\bibitem{rw77}
R.~T. Rockafellar and R.~J.-B. Wets.
\newblock Measures as {L}agrange multipliers in multistage stochastic
  programming.
\newblock {\em J. Math. Anal. Appl.}, 60(2):301--313, 1977.

\bibitem{rw78}
R.~T. Rockafellar and R.~J.-B. Wets.
\newblock The optimal recourse problem in discrete time:
  {$L\sp{1}$}-multipliers for inequality constraints.
\newblock {\em SIAM J. Control Optimization}, 16(1):16--36, 1978.

\bibitem{rw83}
R.~T. Rockafellar and R.~J.-B. Wets.
\newblock Deterministic and stochastic optimization problems of {B}olza type in
  discrete time.
\newblock {\em Stochastics}, 10(3-4):273--312, 1983.

\bibitem{rw98}
R.~T. Rockafellar and R.~J.-B. Wets.
\newblock {\em Variational analysis}, volume 317 of {\em Grundlehren der
  Mathematischen Wissenschaften [Fundamental Principles of Mathematical
  Sciences]}.
\newblock Springer-Verlag, Berlin, 1998.

\bibitem{rog2}
L.~C.~G. Rogers.
\newblock Monte {C}arlo valuation of {A}merican options.
\newblock {\em Math. Finance}, 12(3):271--286, 2002.

\bibitem{sch92}
W.~Schachermayer.
\newblock A {H}ilbert space proof of the fundamental theorem of asset pricing
  in finite discrete time.
\newblock {\em Insurance Math. Econom.}, 11(4):249--257, 1992.

\bibitem{sch4}
W.~Schachermayer.
\newblock The fundamental theorem of asset pricing under proportional
  transaction costs in finite discrete time.
\newblock {\em Math. Finance}, 14(1):19--48, 2004.

\bibitem{sdr9}
A.~Shapiro, D.~Dentcheva, and A.~Ruszczy\'nski.
\newblock {\em Lectures on Stochastic Programming: Modeling and Theory},
  volume~9 of {\em MPS/SIAM Series on Optimization}.
\newblock SIAM, Philadelphia, 2009.

\bibitem{shi96}
A.~N. Shiryaev.
\newblock {\em Probability}, volume~95 of {\em Graduate Texts in Mathematics}.
\newblock Springer-Verlag, New York, second edition, 1996.
\newblock Translated from the first (1980) Russian edition by R. P. Boas.

\bibitem{wet75}
R.~J-B Wets.
\newblock On the relation between stochastic and deterministic optimization.
\newblock In A.~Bensoussan and J.L. Lions, editors, {\em Control Theory,
  Numerical Methods and Computer Systems Modelling}, volume 107 of {\em Lecture
  Notes in Economics and Mathematical Systems}, pages 350--361. Springer, 1975.

\end{thebibliography}

\end{document}